\documentclass[aps,twocolumn,prx,longbibliography]{revtex4-2}
\usepackage{lmodern}
\usepackage{amssymb,amsmath,mathtools}
\usepackage{graphicx}
\usepackage{verbatim}
\usepackage[english]{babel}
\usepackage{xcolor}
\makeatletter
\def\fps@figure{htbp}
\makeatother
\usepackage{xr}
\begin{document}

\title{Concentration dependence of diffusion-limited reaction rates and its consequences}
\author{Sumantra Sarkar}
\affiliation{Center for Nonlinear Studies, Theoretical Division, Los Alamos National Laboratory, Los Alamos, NM 87545, U.S.A. }
%\dates{This manuscript was compiled on \today}

\begin{abstract}
Diffusion-limited association reactions are ubiquitous in nature. They are particularly important for biological reactions, where the reaction rates are often determined by the diffusive transport of the molecules on two-dimensional surfaces, such as the cell membrane. The peculiarities of diffusion on two-dimensional surfaces may lead to nontrivial reaction kinetics, such as concentration dependent rate of association between two molecules. However, traditionally, the kinetics of biomolecular association reactions has been modeled using the law of mass action, which assumes that the rate of reaction is a concentration independent constant. In this paper, using multiscale molecular simulation, we investigate the concentration dependence of diffusion-limited association reactions on 2D surfaces. In particular, we quantify the influence of short-ranged pair interactions on the concentration dependence of the reaction rates and codify it in an empirical law. Using this law in a chemical kinetic model, we find that the the steady state behaviors of simple chemical systems are drastically modified by the presence of concentration dependent rates. In particular, we find that it leads to suppression of intrinsic noise in dimerization reaction and destabilizes robust oscillation in Lotka-Volterra predator-prey systems. In fact, we see a transition from robust to fine-tuned behavior in the latter. In addition, we show that concentration dependent reaction rates arise naturally in stochastic predator-prey systems due to intrinsic noise. We comment on the consequences of these results and discuss their implications in the modeling of complex chemical and biological systems. In particular, we comment on the range of validity of the law of mass action, which is a staple in all theoretical modeling of these systems.
\end{abstract}

%%%%%%%%%%%
\maketitle

\section{Introduction}
Association reactions are a type of elementary reactions, in which two or more reactant molecules form one or more product molecules. Formation of a dimer from two monomers is an example of an association reaction, as is the formation of water from hydrogen and oxygen~\cite{atkinsAtkinsPhysicalChemistry2018}. Due to its elementary nature, association reactions play a central role in many physicochemical processes, including pattern formation~\cite{turingChemicalBasisMorphogenesis1952,crossPatternFormationOutside1993}, aggregation~\cite{barabasi1995fractal}, and cell signaling~\cite{chylekQuantitativeModelingMast2018,harmonTimescaleSeparationPositive2017,chylekPhosphorylationSiteDynamics2014,alonIntroductionSystemsBiology2019,karrWholeCellComputationalModel2012}. A typical association reaction involves two steps~\cite{szaboFirstPassageTime1980,agmonTheoryReversibleDiffusion1990,pastorDiffusionLimitedFirst1996a,takahashiSpatiotemporalCorrelationsCan2010}. In the first step, two molecules are transported near each other through some transport processes. Once the two molecules encounter each other, in the second step, the molecules interact with each other with an intrinsic association rate $\kappa_I$ to form the product molecule. It is usually assumed that the second step is much slower compared to the first step, such that an association reaction occurs after many encounter events. These assumptions have two repercussions: (a) one can assume that the reactant molecules are well mixed, so that the reaction rate is solely determined by the interaction step and (b) the formation of the product molecule follows a Poisson process, such that the rate of association reaction is a time and concentration independent constant~\cite{dibakDiffusioninfluencedReactionRates2019}. When combined, these two observations lead to the celebrated Law of Mass Action (LMA)~\cite{kurtzRelationshipStochasticDeterministic1972, atkinsAtkinsPhysicalChemistry2018}, which states that the propensity of an association reaction is equal to the product of the constant reaction rate, $k_0$, and the mass action $\Phi$, where the latter is the total number of possible reactant pairs. In particular, if we consider the following association reaction
\begin{equation}
  %A + B \underset{k_D}{\stackrel{\kappa_0}{\rightleftharpoons}} AB,
  A + B \underset{}{\stackrel{k_0}{\longrightarrow}} AB,
\end{equation}
then LMA states that the propensity, $r$, of the association reaction is:
\begin{equation}
  r = k_0[A][B] = k_0\Phi,
\end{equation}
where $[.]$ denotes the concentration of a molecule, and $\Phi = [A][B]$ is the mass action.

LMA is strictly valid when the rate $k_0$ is a concentration independent constant. For example, LMA works well for dilute solutions, where this ``rate law" was originally developed. However, its range of applicability did not stay confined within just the purview of dilute solutions. Its simplicity and the ubiquity of association reactions have led to its application in disparate problems~\cite{murray2007mathematical} with varying amount of success. However, it is unclear whether some of the assumptions underlying LMA are still valid for these systems. For example, most cell-signaling reactions occur on two dimensional cell membranes, where, often, the intrinsic reaction rates are higher or comparable to the diffusive encounter rate, such that the reaction kinetics depends crucially on the diffusive transport of the molecules. In particular, the peculiarities of diffusion in two dimension (2D), including nonzero probability of encounter between all particles~\cite{torneyDiffusionLimitedReactionRate1983,yogurtcuTheoryBimolecularAssociation2015}, can result in concentration dependent diffusive encounter rates, which, in turn, results in a concentration dependent $k_0$, which invalidates the application of LMA in such situations~\cite{grimaHowReactionKinetics2006,grimaSystematicInvestigationRate2006}. Furthermore, a fundamental assumption of LMA is that the molecules are essentially point particles that interact with each other only when they collide with each other \cite{vladKineticLawsPhase2009}. In reality, most molecules interact with each other through finite, albeit short-ranged, interactions, which may also influence the diffusion-limited reaction rates. In fact, in such situations also, the diffusion limited reaction rate, $\kappa$, may become concentration dependent~\cite{dibakDiffusioninfluencedReactionRates2019,grimaHowReactionKinetics2006,borosCenterProblemLotka2018} and violate the assumptions of LMA.

 A practical workaround to this challenge is to use an adaptive, concentration dependent, rate of association that varies with time~\cite{szaboFirstPassageTime1980, szaboTheoryDiffusioninfluencedFluorescence1989, torneyDiffusionLimitedReactionRate1983, yogurtcuTheoryBimolecularAssociation2015, gopichMultisiteReversibleAssociation2020}. Use of such adaptive reaction rates in chemical kinetic models drastically improves the prediction of the transient kinetics. Unfortunately, the functional form of the adaptive rates used in these studies are complex and are not immediately conducive to analytical treatments. Hence, in this paper, we offer a complementary empirical formulation of the concentration dependence, which allows us to do analytical computations of the steady state properties. The feasibility of the analytical treatment offers practical advantages to explore the consequences of the concentration dependence for system parameters that are difficult to explore through simulation. 

\begin{figure}
  \centering
  \includegraphics[clip,width=\columnwidth]{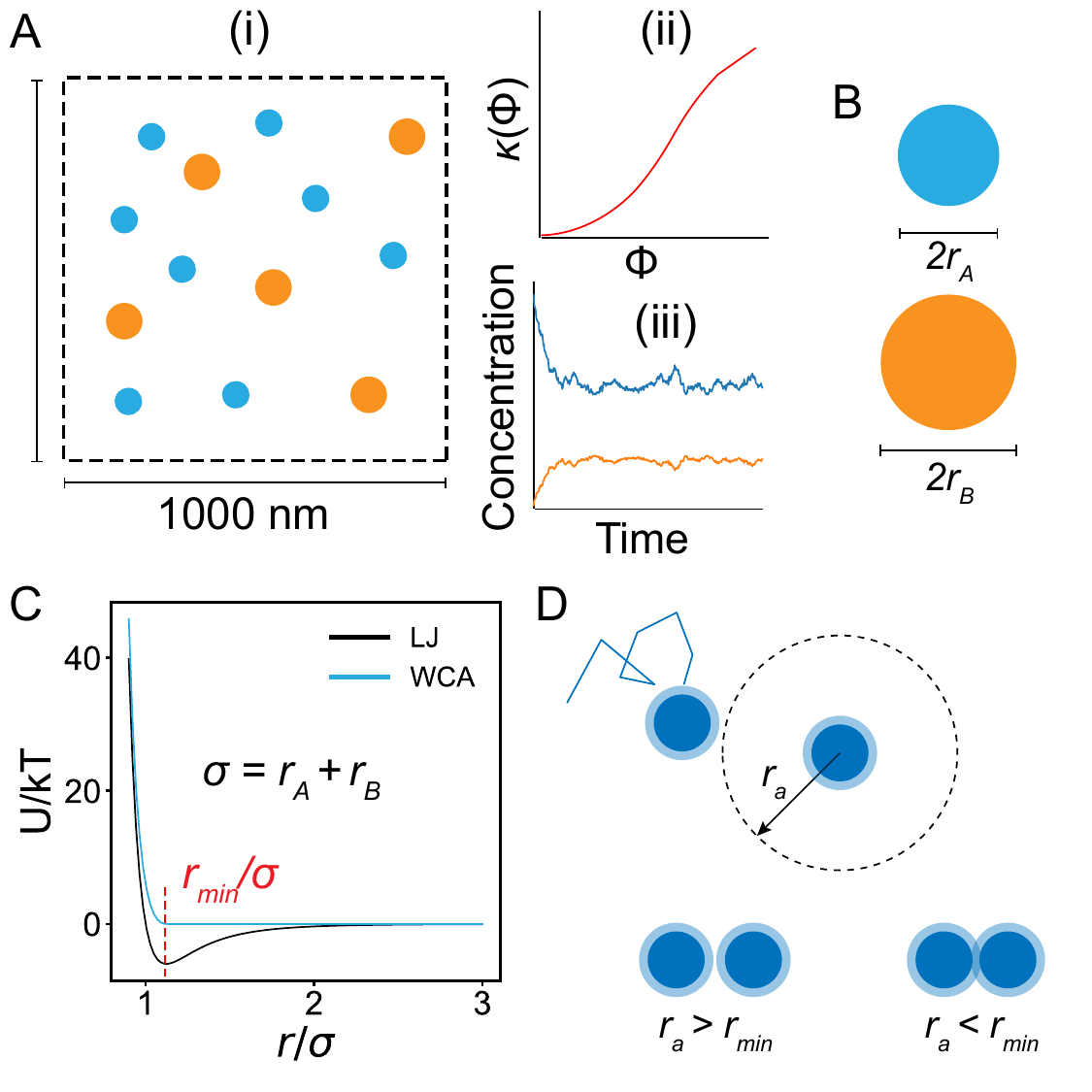}
  \caption{\textbf{Hierarchical multiscale simulation framework:} (A) We combine molecular simulation with chemical kinetic model to study association reaction kinetics at broad spatiotemporal scales. (i) Using BD-GFRD, we study how short-ranged pairwise interaction modifies the rate of association reaction between two molecules on a 2D periodic box of size $1 \mu m \times 1 \mu m$. (ii) The measured rate, $\kappa$ is concentration dependent and its dependence on concentration is represented by the function $\kappa(\Phi)$, where $\Phi$ is the product of the concentrations of the reactants, a.k.a the mass action. (iii) $\kappa(\Phi)$ is used in a spatially homogeneous chemical kinetic model to study concentration variations at timescales not reachable through molecular simulations ($> 1s$). (B) The molecules are represented as disks of radius $r_A$ and $r_B$, and (C) they interact through either Lennard-Jones (LJ) or WCA interaction. (D) The molecules react as soon as they are closer than the reaction radius $r_a$. If $r_a < r_{min}$ (see C and D for definition of $r_{min}$ ), there is strong excluded volume interaction between the reactant molecules.  The light blue region surrounding a molecule denotes the region over which excluded volume interaction is felt by another molecule. More precisely, when two light blue regions touch, the distance between the two molecules is $r_{min}$. }
  \label{fig:setup}
\end{figure}

To theoretically study the concentration dependence of the diffusion-limited reaction rates and to construct an empirical law at concentrations relevant to most applications, one has to simulate the transport and interaction of the molecules in large spatially heterogeneous systems. The main challenge to such line of enquiry is that  diffusion is difficult to investigate through molecular simulation. In this paper, we overcome this challenge by using a recently developed multiscale simulation framework, called the Green's Function Reaction Dynamics with Brownian Dynamics or BD-GFRD~\cite{takahashiSpatiotemporalCorrelationsCan2010,vijaykumarCombiningMolecularDynamics2015,vijaykumarMultiscaleSimulationsAnisotropic2017,sbailoEfficientMultiscaleGreen2017,sokolowskiEGFRDAllDimensions2019}. We combine BD-GFRD with chemical kinetic models to construct a hierarchical multiscale simulation framework (Eq.~\ref{eqn:hierarchical_model}). In the first level of hierarchy of this framework, using BD-GFRD, we measure the concentration dependence of the rates and codify it in a functional form, $\kappa(\Phi)$ (Fig.~\ref{fig:setup}A-i,ii). In the next level, we use $\kappa(\Phi)$ in a chemical kinetic model to study the behavior of the chemical systems at timescales that are not reachable through BD-GFRD(Fig.~\ref{fig:setup}A-iii). We ask, can we still use LMA even when the reaction rates are concentration dependent? In the two model systems that we study here, we find that steady state properties obtained from concentration dependent rates are qualitatively different from LMA, but under some special conditions this difference is negligible. Because chemical kinetic models are staple in many lines of scientific enquiry~\cite{palssonSystemsBiologyProperties2006}, our results may provide useful guidelines and design principles for these models.

\begin{figure*}
\begin{eqnarray}
  \text{Molecular Simulation} \underset{}{\stackrel{\text{generates}}{\longrightarrow}} \kappa(\Phi)\underset{}{\stackrel{\text{is used to construct}}{\longrightarrow}\text{Chemical Kinetic Model}\underset{}{\stackrel{\text{is used to compute}} {\longrightarrow}}\text{Steady state behavior}}
  \label{eqn:hierarchical_model}
\end{eqnarray}
\end{figure*}

%Using GFRD, we study the concentration dependence of diffusion-limited reaction rates on 2D periodic plane at biologically relevant concentrations and timescales. To understand the consequences of the concentration dependent rates for timescales beyond the reach of molecular simulations,

%\section{Results}
\section{Concentration dependence of the diffusive collision rate}

\subsection{Molecular simulation of dimerization reactions}

\begin{figure*}
  \centering
  \includegraphics[width=0.95\textwidth]{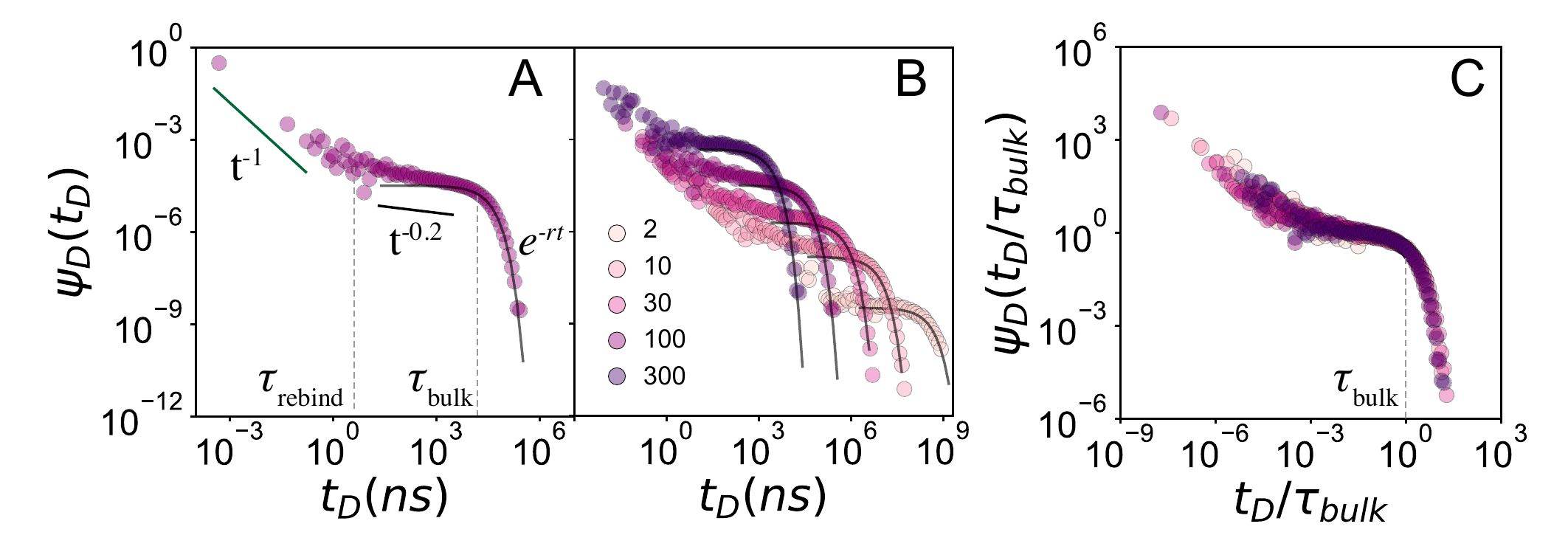}
  \caption{(A) The dimerization time interval distribution, $\psi_D(t_D)$. There are three distinct regions separated by two timescale $\tau_{rebind}$ and $\tau_{bulk}$. Reactions occurring at intervals $t_D > \tau_{bulk}$ are described by a single rate and their propensities can be computed by fitting the exponential tail. The $t^{-1}$ distribution for $t_D < \tau_{rebind}$ arises because the dimerization events are time-correlated due to rebinding events. (B) $\psi_D(t_D)$ for different monomer concentrations (shown in the legend with unit $\mu m^{-2}$). The size of the region between $\tau_{rebind} < t_D < \tau_{bulk}$, i.e. the region with $t^{-0.2}$ scaling, decreases with decreasing monomer concentrations. In fact, at monomer concentration of $10\mu m^{-2}$, this region is almost nonexistent, as illustrated by the exponential fit. The origin of the $t^{-0.2}$ scaling remains unclear (see text). (C) Rescaling $t_D$ by $\tau_{bulk}$ collapses $\psi_D(t_D)$ for different concentrations onto a single master curve. 
  }

  %For timescales below $\tau_{rebind}$, the distribution scales as $t^{-1}$. For timescales above the diffusive encounter time $\tau_{bulk}$, the distribution decays exponentially. For intermediate times, the distribution scales as $t^{-0.2}$. We measure the reaction rates by fitting an exponential function for $t_D > \tau_{bulk}$. The reactions at this timescales are slower than diffusive encounter rate and well-mixed approximation holds.}
  \label{fig:tD}
\end{figure*}

To study how short-ranged pairwise interaction leads to concentration dependent reaction rates, we use a hierarchical simulation framework, which interfaces a recently developed multiscale molecular simulation method called BD-GFRD~\cite{takahashiSpatiotemporalCorrelationsCan2010,vijaykumarCombiningMolecularDynamics2015,vijaykumarMultiscaleSimulationsAnisotropic2017,sbailoEfficientMultiscaleGreen2017,sokolowskiEGFRDAllDimensions2019} with chemical kinetic models to study association reaction kinetics at wide spatiotemporal scales and broad concentration levels (Fig.~\ref{fig:setup}A-D and Eq.~\ref{eqn:hierarchical_model}). Please check appendix B to find the detailed description of the methods. 

To measure the impact of short-ranged interactions (Fig.~\ref{fig:setup}C) on diffusion-limited reaction rates, we investigate the concentration dependence of the rates of the following reactions:
\begin{eqnarray}
% A + A \underset{k_d}{\stackrel{\kappa}{\rightleftharpoons}} A_2 \\
A + A &\rightarrow& A_2 \label{eqn:homodimer}\\
A + B &\rightarrow& AB
\end{eqnarray}

We measure the diffusive encounter rates of these reactions in an ensemble in which the number of each species of molecules is conserved. Using this set-up in the BD-GFRD~\cite{takahashiSpatiotemporalCorrelationsCan2010,vijaykumarCombiningMolecularDynamics2015,vijaykumarMultiscaleSimulationsAnisotropic2017,sbailoEfficientMultiscaleGreen2017,sokolowskiEGFRDAllDimensions2019} simulations (see appendix B), we measured the reaction rates by varying the concentration of the monomers in the range of $2-300 /\mu m^2$. These concentrations are similar to the concentration of integral and peripheral membrane proteins on cell-membranes.

We measured probability distribution, $\psi_{D}\left( t_{D} \right)$, of the time interval, $t_D$, between two consecutive homodimerization reactions. It has three distinct regimes separated by two timescales: the rebinding timescale $\tau_{rebind}$ and the timescale above which the reaction events are described by an exponential decay: $\tau_{bulk}$ (Fig.~\ref{fig:tD} A). For $t_D < \tau_{rebind}$, $\psi_{D}\left( t_{D} \right)$ decays as $t_D^{-1}$. Such dependence occurs due to rapid rebinding of the monomers to form dimers~\cite{takahashiSpatiotemporalCorrelationsCan2010}. For $\tau_{rebind} < t_D < \tau_{bulk}$, $\psi_{D}\left( t_{D} \right)$ decays as $t_D^{-0.2}$.  We find that this region is present in all the monomer concentrations considered here (Fig.~\ref{fig:tD} B). As the concentration of the monomers decreases, the width of this region also decreases, which suggests that this unusual scaling results from reactions that occur before the particles have lost their memory of the previous encounter. Beyond this observation, we do not understand this scaling well and plan to investigate it in a future paper. For timescales above $\tau_{bulk}$,  $\psi_{D}(t_{D})$ decays exponentially (Fig.~\ref{fig:tD}A). In particular, we find that rescaling the time by $\tau_{bulk}$ collapses all distributions on to a single master curve (Fig.~\ref{fig:tD}C), which implies that the propensity of the reactions $r$ is inversely proportional to $\tau_{bulk}$. $\tau_{bulk}$ depends on the concentration of the monomers (Fig.~\ref{fig:SI-tau-bulk}) and at low $\Phi$ is given by:
\begin{equation}
  \tau_{bulk} = \frac{\log(L/a)}{8\pi D_{A}\Phi},
  \label{eq:tau_{bulk}}
\end{equation}
where $L = 1000 nm$ is the system size, $a = 2r_A$, $D_{A} = 1 \mu m^2s^{-1}$ is the diffusion constant of $A$, and $\Phi = [A]([A]-1)/2$~\cite{yogurtcuTheoryBimolecularAssociation2015}. We observe the same behavior for the heterodimerization reactions as well (not shown). Crucially, we observe identical $\psi_D(t_D)$ in a set of simulations in which $[A] + 2[A_2]$ was kept fixed, but concentrations of $[A]$ and $[A_2]$ could vary, which consolidates our observation (Fig.~\ref{fig:SI-dimer-canonical}). One should note that, although $\psi_{D}(t_{D})$ is described by an exponential decay in this regime, it does not imply that this process is Markovian. In 2D diffusion, due to the finite probability of reencounter between two reactants, there is not a single well-defined rate of reaction for all concentrations. Instead, the rate depends on the concentrations of the reactants, which can be used in a manner similar to the Markovian rate constant~\cite{szaboFirstPassageTime1980, szaboTheoryDiffusioninfluencedFluorescence1989, torneyDiffusionLimitedReactionRate1983, yogurtcuTheoryBimolecularAssociation2015, gopichMultisiteReversibleAssociation2020}.

\subsection{Diffusion-limited bulk dimerization rate}
\begin{figure}
  \centering
  \includegraphics[clip,width=0.7\columnwidth]{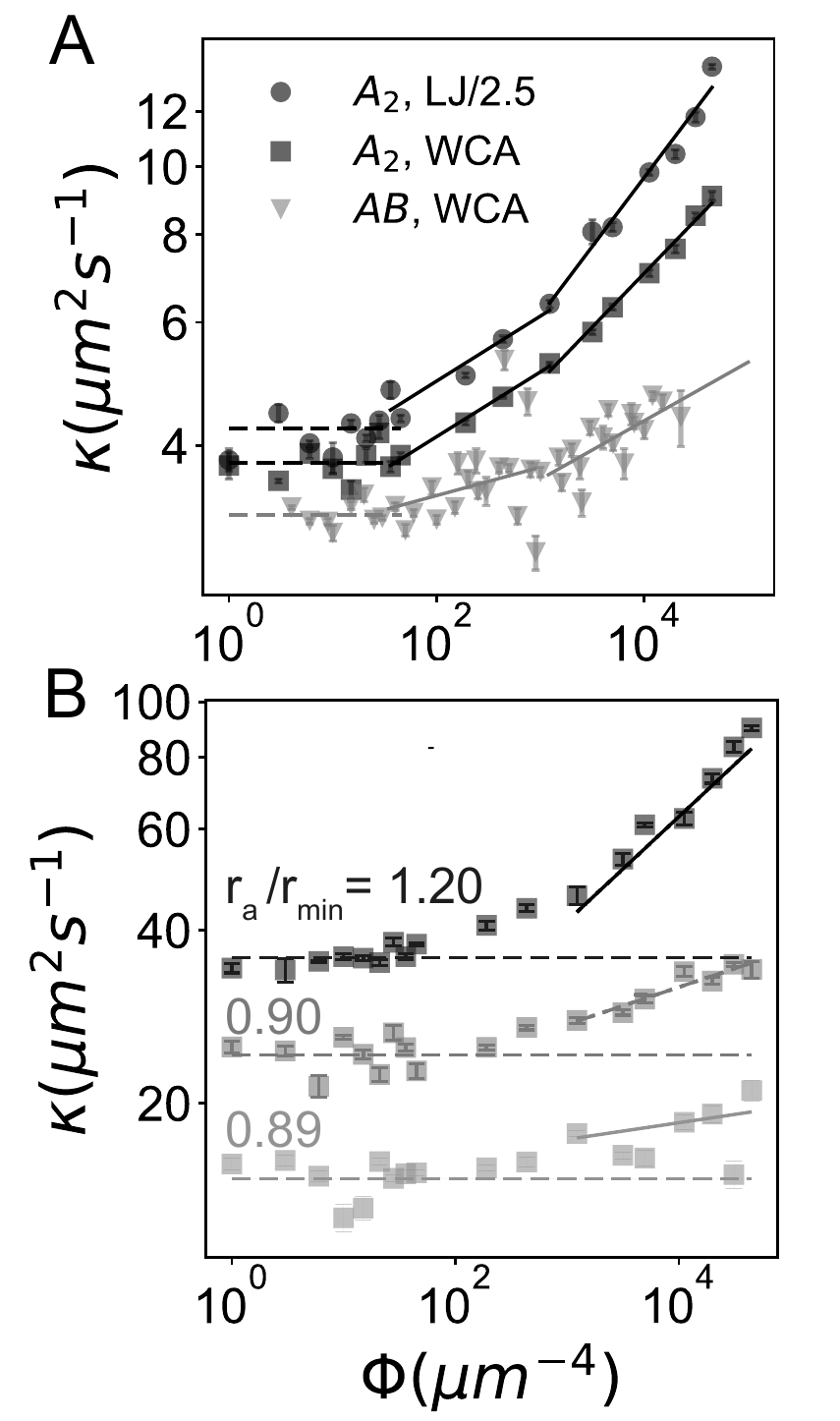}
  \caption{\textbf{Concentration dependent diffusion-limited reaction rate:} (A) The rates of association reaction, $\kappa$ depends on the mass action $\Phi$. The functional form of $\kappa(\Phi)$ is independent of the interaction potential, but depends on the association reaction. The solid lines are guide for the eye, whereas the dashed lines show the value of $\kappa_0 = \lim_{\Phi\rightarrow 0} \kappa(\Phi)$. (B) $\kappa(\Phi)$ depends on the reaction radius $r_a$. The lower the $r_a$, the more slowly $\kappa(\Phi)$ increases with $\Phi$. To compute these results we used $D = 10\mu m^2/s$. The change in diffusion constant does not change the result qualitatively(Fig.~\ref{fig:SI-D-dependence}).}

  \label{fig:kappa}
\end{figure}

The propensities, $r$, of the dimerization reactions separated by $t_D > \tau_{bulk}$ can be computed by fitting the tail of $\psi_D(t_D)$ with an exponential function, from which we can determine $\kappa(\Phi)$, the ``bulk" concentration dependent diffusion-limited reaction rate, using the following formula:
\begin{equation}
  r = \kappa(\Phi)\Phi
  \label{eqn:kappa_phi}
\end{equation}
For LMA to hold, $\kappa(\Phi)$ should be independent of $\Phi$. However, we find that $\kappa$ varies with $\Phi$ for all pair interactions considered here (Fig.~\ref{fig:kappa}A). Irrespective of the type of interaction potential, the functional form of $\kappa(\Phi)$ is nearly identical for $A_2$, except its value near $\Phi = 0$. The heterodimerization reaction, i.e. the formation of $AB$, also has similar concentration dependence. However, $\kappa(\Phi)$ varies weakly compared to $A_2$. Finally, $\kappa(\Phi)$ also depends on the reaction radius $r_a$~\cite{doiTheoryDiffusioncontrolledReaction1975,doiTheoryDiffusioncontrolledReaction1975a} (Fig.~\ref{fig:setup}D). We found that the lower the $r_{a}$, the more slowly $ \kappa(\Phi) $ increases with $\Phi$. In fact, when $r_{a} = 0.89r_{\min} \approx \sigma$, $\kappa(\Phi)$ does not change appreciably for the $\Phi$ values considered in our simulations (Fig.~\ref{fig:kappa}B). In fact, for all the parameters considered here, as $\Phi\rightarrow0$, $\kappa(\Phi)$ converges to a concentration independent value $\kappa_0$, which is consistent with previous works~\cite{yogurtcuTheoryBimolecularAssociation2015,dibakDiffusioninfluencedReactionRates2019,sarkarPresenceAbsenceRas2020,ngoHowAnionicLipids2020}.

\subsection{Empirical rate law}
\begin{figure}
  \centering
  \includegraphics[clip,width=0.7\columnwidth]{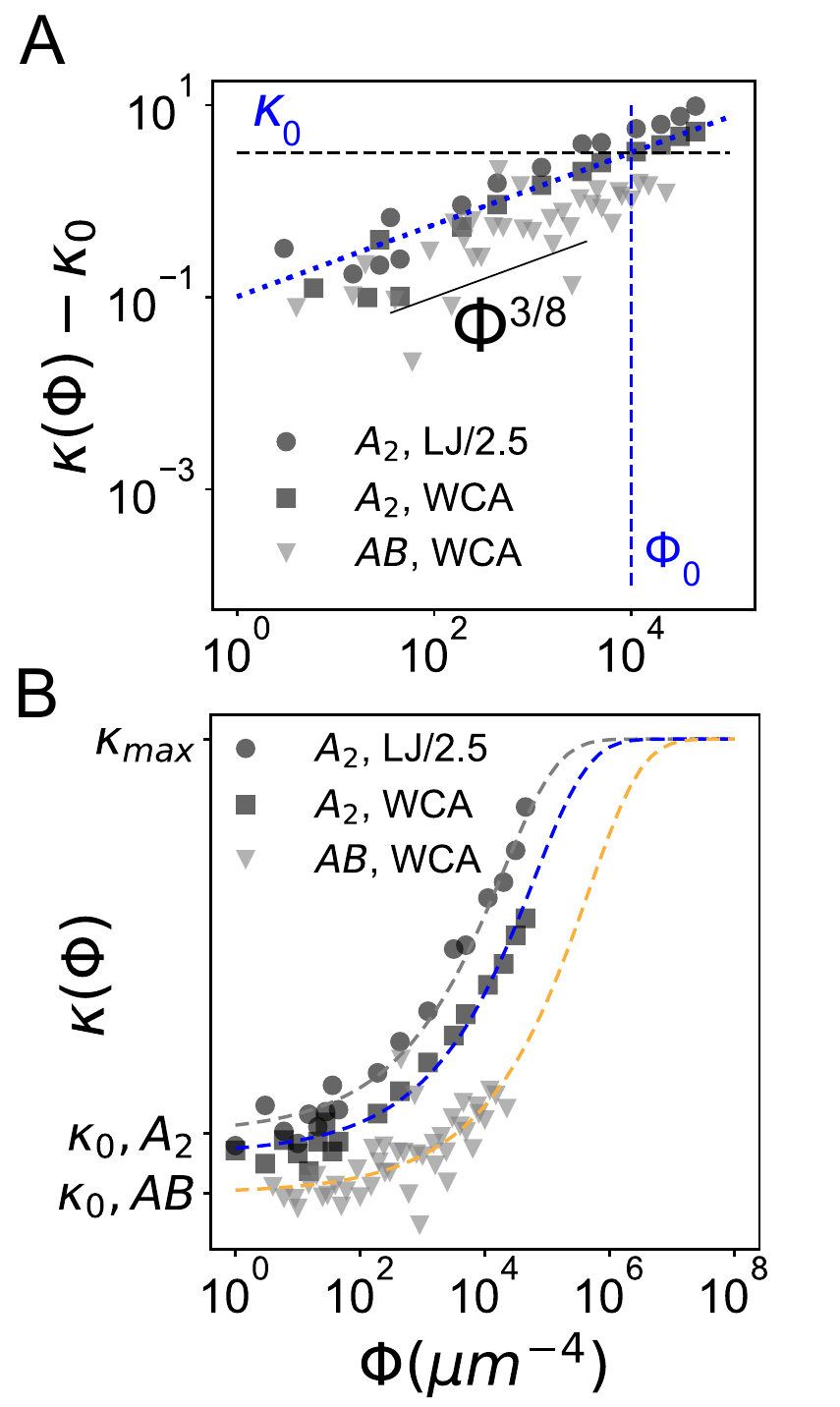}
  \caption{\textbf{Empirical rate law:} (A) $\kappa(\Phi) - \kappa_0$ increases as $\kappa_1\Phi^{\epsilon}$, with $\epsilon = 3/8$, for the $\Phi$ values considered in our simulations (blue dotted line). $\kappa_0$ is the concentration independent rate and $\Phi_0$ is the mass-action flux where $\kappa_1\Phi_0^{3/8} = \kappa_0$. (B) This scaling is used to postulate a sigmoidal $\kappa(\Phi)$. The parameters of the sigmoidal functions are $\kappa_0$, $\kappa_{max}$ and $\Phi_0$ (see text for details). We use the postulated sigmoidal function as an empirical rate law for $\Phi$ values both encountered and not encountered in the molecular simulation. To fit these curves we have used: $\kappa_0 = 4.2 \mu m^2 s^{-1}$ and $\Phi_0 = 5\times 10^3$ for $A_2$ in the presence of LJ interaction (dashed grey); $\kappa_0 = 3.7 \mu m^2 s^{-1}$ and $\Phi_0 = 1\times 10^4$ for $A_2$ in the presence of WCA interaction (dashed blue); and $\kappa_0 = 3.2 \mu m^2 s^{-1}$ and $\Phi_0 = 1\times 10^5$ for $AB$ in the presence of WCA interactions (dashed orange). The values of $\kappa_0$ were obtained by averaging $\kappa(\Phi)$ values for $\Phi < 30$ for $D = 1\mu m^2/s$ and $r_a = 1.2 r_{min}$ (Fig.~\ref{fig:kappa}A).}
  
  \label{fig:empirical}
\end{figure}

The preceding discussion implies that, in general, $\kappa(\Phi)$ can be expressed in terms of $\kappa_0$, the value of $\kappa(\Phi)$ as $\Phi \rightarrow 0$, and a function of the mass action $\Phi$. The latter can be obtained by measuring the variation of $\kappa(\Phi) - \kappa_0$ with $\Phi$. Remarkably, we find that $\kappa(\Phi) - \kappa_0$ displays universal variation across different reactions studied here. In particular, for the concentrations of the monomers studied here, we find that $\kappa(\Phi) - \kappa_0 \sim \Phi^{\epsilon}$, such that $\kappa(\Phi) \approx \kappa_0 + \kappa_1\Phi^{\epsilon}$,  where $\epsilon = 3/8 = 0.375$ (Fig.~\ref{fig:empirical}A). Furthermore, we find that there is a mass-action value $\Phi_0$ such that $\kappa_1\Phi_0^{\epsilon} = \kappa_0$, i.e., $\Phi_0$ measures how quickly  a reaction deviates from the concentration independent behavior; the larger the value of $\Phi_0$, the slower is the deviation. Combining the above definitions, $\kappa(\Phi)$ can be succinctly written as: 

\begin{equation}
\kappa(\Phi) \approx \kappa_0\left( 1 + \left(\frac{\Phi}{\Phi_0}\right)^{\epsilon}\right)\label{eqn:approx_kappa}
\end{equation}

This form of $\kappa(\Phi)$ is valid for very low packing fractions of the reactants. For example, the packing fraction of the molecules at $300 /\mu m^2$, the highest concentration of molecules studied here, is approximately 0.3\%, which is much lower than the packing fraction of molecules in a cellular environment (up to 40\%~\cite{grimaSystematicInvestigationRate2006}). Therefore, we expect that this approximate form of $\kappa(\Phi)$ will not hold at such high packing fractions. Instead, because of the increase in viscosity observed at high packing fractions, we expect the diffusive encounter rate $\kappa(\Phi)$ to saturate to a maximum value, $\kappa_{max}$~\cite{bulowDynamicClusterFormation2019}. It is possible to obtain the value of $\kappa_{max}$ by doing direct simulation of the molecules at high density. Although we plan to study such a system, we do not attempt to do it in this paper. Instead, we use Smoluchowski's theory of diffusion limited reactions~\cite{yogurtcuTheoryBimolecularAssociation2015} to obtain $\kappa_{max}$, which for the parameters of our system is approximately $ 20 \mu m^2/s$ (appendix E). Furthermore, because $\kappa(\Phi)$ is a convex function for small $\Phi$ and saturates to a maximum value at higher $\Phi$ values, we expect its most general form to be sigmoidal. Therefore, we postulate the following empirical law that describes $\kappa(\Phi)$ for concentrations beyond the values simulated in this paper:

\begin{equation}
  \kappa(\Phi) = \frac{\kappa_0\kappa_{max}}{\kappa_0 + (\kappa_{max}-\kappa_0)\exp\left[-\left(\frac{\Phi}{\Phi_0}\right)^{\epsilon}\right]}. \label{eq:empirical_law}
\end{equation}

It is easy to check that this sigmoidal $\kappa(\Phi)$ has the desired behavior for different limits of $\Phi$. Also, it fits the observed rates well (Fig.~\ref{fig:empirical}B).    In fact, the functional form that we obtain here is very similar to what has been described earlier~\cite{yogurtcuTheoryBimolecularAssociation2015}. The advantage of our approach is that the concentration dependence of the encounter rate is entirely codified by the flux $\Phi$, whereas everything else are concentration independent parameters, and can, in principle, be obtained by using self-consistent theories of association reaction kinetics~\cite{szaboFirstPassageTime1980, szaboTheoryDiffusioninfluencedFluorescence1989, torneyDiffusionLimitedReactionRate1983, yogurtcuTheoryBimolecularAssociation2015, gopichMultisiteReversibleAssociation2020}. Empirical law in this simple form is not only useful as an input to the numerical methods, such as Gillespie algorithm, it is also conducive to analytical treatments, such as the calculation of the steady state concentrations and their stabilities, as described in the next section. 

In the rest of this paper, we study the consequences of this sigmoidal rate law using chemical kinetic models (see appendix B). For the sake of brevity, we shall refer to this rate law as the ``law of concentration dependent rates" or LCDR, \textit{ala} LMA. We study two model chemical systems using both LMA and LCDR and compare the steady state behavior of these models under the two different rate laws. We should point out that ours is by no means the first such attempt at using chemical kinetic model to study the long-term behavior of chemical systems. In fact, there has been several papers that have pointed out that chemical kinetic equations are not accurate at low concentrations~\cite{yogurtcuTheoryBimolecularAssociation2015, gopichMultisiteReversibleAssociation2020}. However, in spite of this limitation, the chemical kinetic models with concentration dependent rate constants have been shown to reproduce kinetics of association reactions with nearly exact transient kinetics~\cite{yogurtcuTheoryBimolecularAssociation2015,gopichMultisiteReversibleAssociation2020}. Therefore, we are justified in using such an approach to study the consequences of the concentration dependent rates.

\section{Consequences of concentration dependent rates}

\subsection{Suppression of intrinsic noise}
\begin{figure}
  \centering
  \includegraphics[clip,width=0.7\columnwidth]{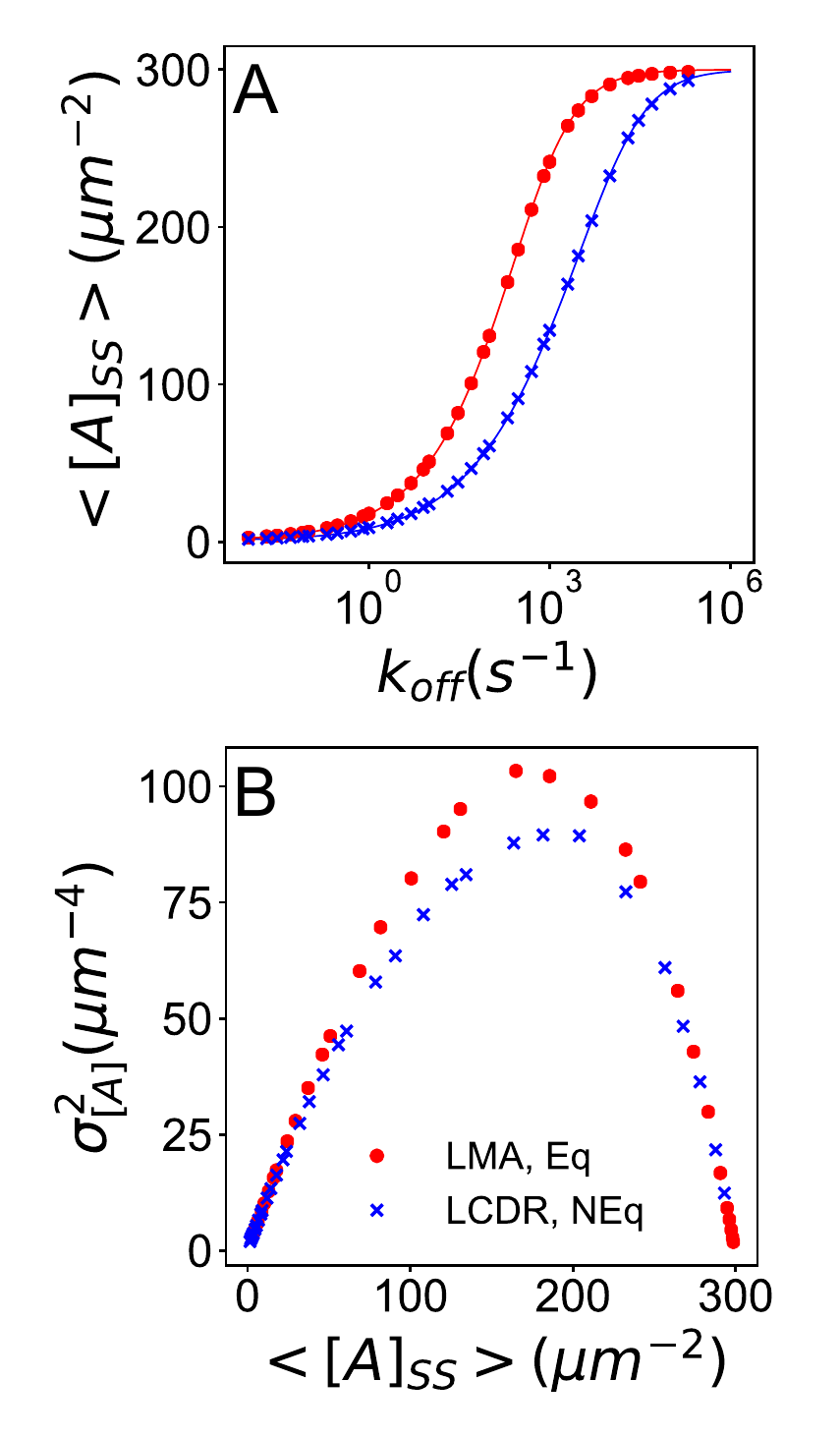}
  \caption{\textbf{Suppression of noise in dimerization reaction:} (A) The mean steady state concentration of the monomers, $<[A]_{ss}>$ as a function of dissociation rate of the dimer $k_{off}$. To compare LMA with LCDR, we have assumed that $k_{on} = \kappa_0 = 4 \mu m^2 s^{-1}$ for LMA, because then $k_{off}$ is identical in both equilibrium and nonequilibrium simulations . For LCDR, we have used $\kappa_0 = 4 \mu m^2 s^{-1}$, $\kappa_{max} = 20 \mu m^{2}s^{-1}$, and $\Phi_0 = 5\times 10^3 \mu m^{-4}$. The simulation results (see legend in B) match exactly with the analytically calculated values (lines). (B) The variance of monomer concentrations, $\sigma^2_{[A]}$ vs $<[A]_{ss}>$. The variance is identical between LMA and LCDR at chemical equilibrium (LCDR not shown). However, in nonequilibrium steady states, the variance is lower for LCDR than LMA for same $<[A]_{ss}>$ at intermediate concentrations. For both simulations, we have used $[A] + 2[A_2] = 300 \mu m^{-2}$.  }
  \label{fig:dimer}
\end{figure}

We studied the consequences of the concentration dependent reaction rates on the simplest association reaction possible: the homodimerization reaction.
\begin{equation}
A + A \underset{k_{off}}{\stackrel{k_{on}}{\rightleftharpoons}} A_2, 
\end{equation}
where $k_{on} =  \kappa(\Phi)$ is the association rate; $\kappa_{max} = 20 \mu m^2 s^{-1}$ and $k_{off}$ is the dissociation rate of the dimer. 
 
We study the steady state properties of this reaction using a chemical kinetic model. To do so, we measure the propensities of the association reactions using Eq.~\ref{eqn:kappa_phi}, where $\Phi = [A]([A] - 1)/2$ is the mass action and $\kappa(\Phi)$ is given by Eq.~\ref{eq:empirical_law}. We study the steady state properties of this system by varying the dissociation rate, $k_{off}$. To obtain equilibrium steady states, the rates have to obey detailed balance, i.e., $k_{off}$  has to be equal to $k_{on}k_D$, where $k_D$ is the equilibrium constant. To compute the mean equilibrium concentration and fluctuation at different $k_D$ values, we simulated the dimerization reactions using Gillespie algorithm. The mean equilibrium concentrations, $<[A]_{ss}>$, and the variance of the fluctuations in the steady state, $\sigma^2_{[A]}$,  are identical between LMA and LCDR (Fig.~\ref{fig:dimer}A-B), which is what we expect. To verify the correctness of the simulated results, we also calculated the steady state concentrations analytically for both LMA and LCDR (see appendix C). We found that the mean monomer concentration $<[A]_{ss}>$ obtained from analytical calculation and simulation are identical (Fig.~\ref{fig:dimer}A).

To study the same quantities in a nonequilibrium steady state, we use $k_{off} = k_D$. For LMA, for which $\kappa(\Phi)$ is a concentration independent constant, this choice of the dissociation rate does not break detailed balance and we reproduce the equilibrium mean concentration and fluctuation profile. In contrast, for LCDR, this functional form of $k_{off}$ is sufficient to break detailed balance and obtain a nonequilibrium steady state. In addition, we discover that for same steady state concentrations, the variance of the steady state fluctuations are smaller in the nonequilibrium steady state of LCDR than in equilibrium state under both LMA and LCDR kinetics (Fig.~\ref{fig:dimer}B). This is a remarkable result and can be explained through a simple analogy. To do so, we first write $\kappa(\Phi)$ in its approximate form for $\Phi$ near $\Phi_0$:
\begin{equation}
  \kappa(\Phi) \approx \kappa_0 + \kappa_1\Phi^{\epsilon}.
  \label{eq:kappa_spring}
\end{equation}
Written in this form $r = \kappa(\Phi)\Phi$ can be interpreted as a cost function, which computes the cost associated with maintaining the system at a particular mass action value $\Phi$. LMA, for which $\kappa_1$ is zero, imposes a penalty $\kappa_0\Delta\Phi$ for a concentration fluctuation that changes $\Phi$ to $\Phi + \Delta\Phi$. On the other hand, LCDR imposes an additional penalty of $(1+\epsilon)\kappa_1\Delta\Phi\Phi^{\epsilon}$.  In equilibrium, this additional cost is balanced by the reverse reaction, whose propensity is equal to $\kappa(\Phi)k_D[A_2]$. Hence, the fluctuation profiles are identical between LMA and LCDR in equilibrium. In contrast, in the nonequilibrium state, at a given value of $\Phi$, the reverse reaction imposes a constant cost, given by $k_D[A_2]$. Therefore, for all $\Phi > 0$, fluctuations cost more in LCDR than in LMA. In fact, for LCDR, the cost of fluctuation increases with increasing $\Phi$, but, for LMA, the cost of fluctuation is independent of $\Phi$. That is, the difference in variance should increase with increasing monomer concentration, which is what we observe in Fig.~\ref{fig:dimer} until $<[A]_{ss}> \approx 150 \mu m^{-2}$, beyond which the finiteness of the system becomes important.

We must stress that the attainment of the nonequilibrium steady state was possible due to the concentration dependence of the reaction rates (see appendix C). As we have shown, in the absence of the concentration dependence, as in LMA, the system always reaches chemical equilibrium, unless the reverse reaction is time-dependent. However, the reader should note that our study of the effect of concentration dependent reaction rate on the  dimerization reaction is a toy model through which we have decided to expose the impact of the concentration dependent rate. To assess the practical relevance of our results, we still need to verify whether, in real systems, detailed balance is broken in the manner described in this paper. However, as the preceding analogy shows, the underlying reason for the noise suppression is the additional cost of fluctuation due to the concentration dependence, which does not depend on the choice of the nonequilibrium steady state.  Therefore, the suppression of intrinsic noise by concentration dependent rates may be a useful strategy for controlling noise. 

Control of noise is an important aspect of many biological functions~\cite{ozbudak2002regulation,thattai2001intrinsic,sanchez2013regulation,grimaNoiseInducedBreakdownMichaelisMenten2009,grimaIntrinsicBiochemicalNoise2010,mayawalaSpatialModelingDimerization2006}. It is possible that cellular systems may control noise by modulating the local concentration of reactants on 2D surfaces, such as membranes, which can increase or decrease the amount of noise. Indeed, liquid-liquid phase separation of signaling proteins has been postulated to reduce noise in the processing of the signals~\cite{huangMolecularAssemblyPhase2019, klosin2020phase}. Therefore, we anticipate our results will provide important insights in deciphering the nature of noise control through such biological mechanisms.

%We observe that the steady state is approached more rapidly in LCDR than in LMA, because the propensities of the association reaction is higher in LCDR due to the concentration dependent $\kappa(\Phi)$.  Both show that $<[A]_{ss}>$ increases more slowly with $k_D$ for LCDR than for LMA. In particular, the stronger the deviation from LMA, i.e. the lower the $\Phi_0$, the slower does the $<[A]_{ss}>$ grow with $k_D$ (Fig.~\ref{fig:SI_ssn}). Furthermore, we find that for same $<[A]_{ss}>$, the variance of the steady state concentration, $\sigma^2_{[A]}$, is lower for LCDR than LMA (Fig.~\ref{fig:dimer}).

\subsection{Chemical oscillation becomes fine-tuned}
\begin{figure}
  \centering
  \includegraphics[clip,width=\columnwidth]{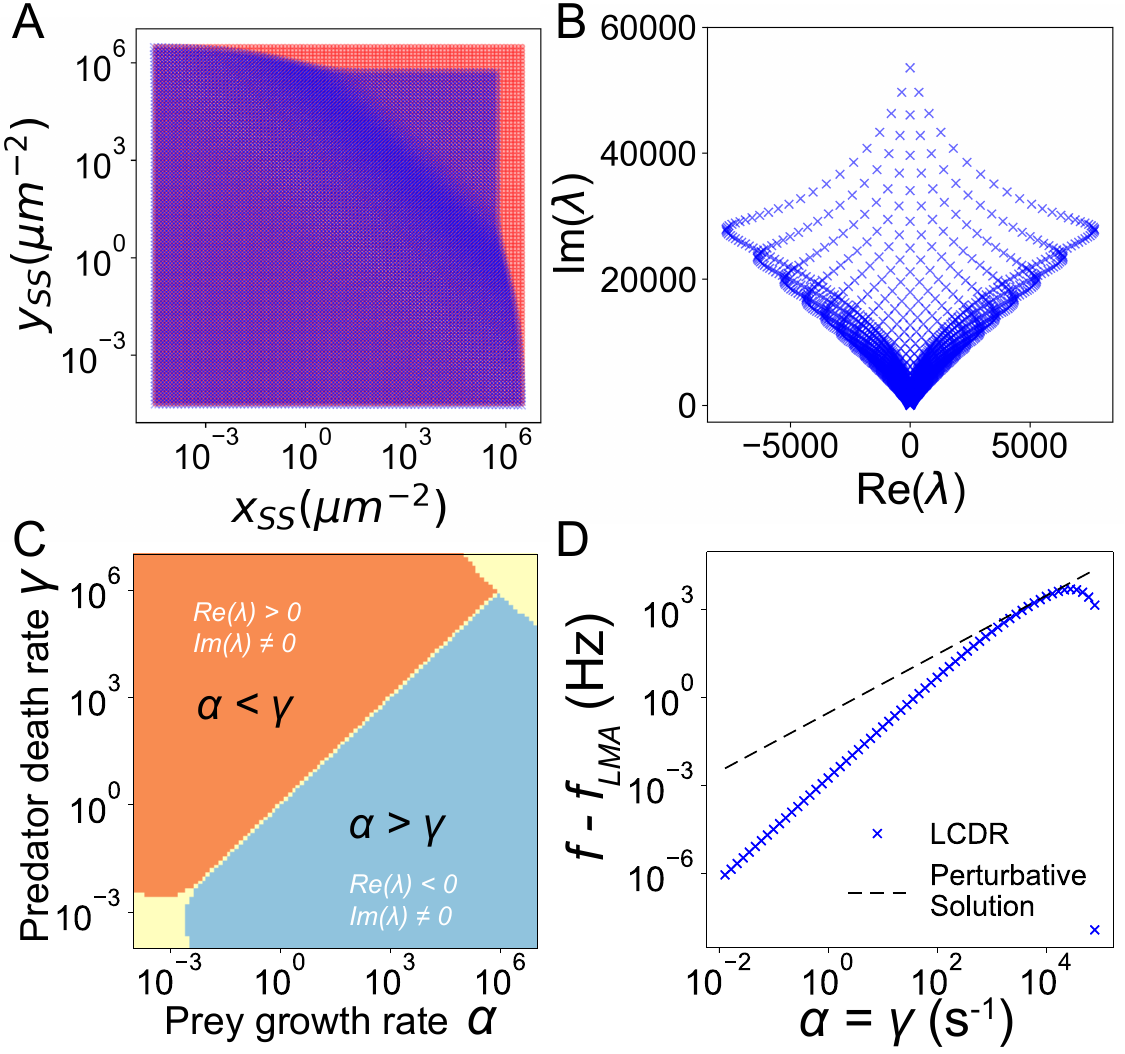}
  \caption{\textbf{Fine tuning of oscillation:} Solution of Lotka-Voterra predator-prey model using LMA and LCDR leads to very different steady state properties. (A) The steady state prey ($x_{ss}$) vs predator ($y_{ss}$) concentration when LMA is used (red dots) and when LCDR is used (blue crosses). The compression of phase space shows that LCDR constraints large variation in concentration, which is consistent with Fig.~\ref{fig:dimer}. (B) The real part of the eigenvalues of the Jacobian vs their imaginary parts show that most steady state solutions are non-oscillatory when LCDR is used. (C) The phase diagram of LV predator-prey system. Under LCDR, sustained oscillation of concentration(pale yellow) happens only when $\alpha = \gamma$ for a wide range of the parameter values, which is a drastic reduction from LMA, where oscillation occurs for all $\alpha$ and $\gamma$. (D) The deviation in the frequency of oscillation in LCDR from that predicted by LMA, $f - f_{LMA}$, vs $\alpha$, in the parameter range where oscillation happens only when $\alpha = \gamma$. Perturbation theory (see appendix D) correctly captures the difference in a small range of $\alpha$ (dashed black line). }
  \label{fig:LV}
\end{figure}

\subsubsection{Lotka-Volterra model in the presence of concentration dependent rates}
Next, we consider the consequences of the pairwise interaction on the behavior of chemical oscillators. We study the Lotka-Volterra (LV) predator-prey equations~\cite{murray2007mathematical}, which describes the dynamics of the following set of reactions:
\begin{eqnarray}
X &\rightarrow& 2X \nonumber\\
X + Y &\rightarrow& 2Y \label{Eq:LV-reactions}\\
Y &\rightarrow& \phi \nonumber,
\end{eqnarray}
where $X$ is the prey and $Y$ is the predator, and $\phi$ represents death. The most general LV equation can be written in the following way:
\begin{eqnarray}
\frac{dx}{dt} &=& \alpha x - \beta(x,y)xy\nonumber\\
\frac{dy}{dt} &=& \beta(x,y)xy - \gamma y,
\label{Eq:LV-ODE}
\end{eqnarray}
which admits two steady state solutions: $x = y = 0$ and $x = \frac{\gamma}{\beta(x,y)}, y = \frac{\alpha}{\beta(x,y)}$, where $x$ and $y$ are the concentrations of $X$ and $Y$, and $\beta(x,y) = \beta(xy)$ is the concentration dependent predation rate. We assume that $\beta(xy)$ has the same functional form as $\kappa(\Phi)$ with $\Phi = xy$ and $\Phi_0 = 10^5 \mu m^{-4}$.The concentration of the nonzero steady state depends on the kinetics used. For LMA, i.e. $\beta(x,y) = \beta_0$, the steady state concentrations increase linearly with $\alpha$ and $\gamma$. In contrast, for LCDR, the steady state concentrations are highly nonlinear when either $\alpha$ or $\gamma$ is large, but linear behavior is restored when both $\alpha$ and $\gamma$ are small or when the steady state mass action $\Phi_{ss} >> \Phi_0$ (Fig.~\ref{fig:LV}A) . %As a result, the phase space spanned by the steady state concentrations is highly compressed in LCDR, compared to LMA.

While the $x = y = 0$ steady state is a saddle point for both kinetics, the stability of the nonzero steady state depends on the underlying kinetics. For LMA, i.e. $\beta(x,y) = \beta_0$, the steady state is oscillatory with frequency $\sqrt{\alpha\gamma}$ for all values of $\alpha$ and $\gamma$. That is, the oscillation of the prey and the predator population is robust to the variation of the parameters. In LCDR, we do see similar behavior when $\alpha, \gamma < 10^{-2}$ or when $\alpha, \gamma > 10^{5}$, where $\kappa(\Phi)$ is effectively $\Phi$ independent. However, outside these parameter ranges, the stability of the steady state depends on $\alpha$ and $\gamma$, as can be seen from the eigenvalues at different parameter values(Fig.~\ref{fig:LV}B). In particular, when $\alpha > \gamma$, the steady state is a stable fixed point. When $\alpha < \gamma$, the steady state is an unstable fixed point and admits exponential rise of the peak predator population and exponential fall of the prey population. Finally, only when $\alpha = \gamma$, the steady state solution is periodic (Fig.~\ref{fig:LV}C). Therefore, in LCDR, the parameters have to be fine tuned to achieve sustained oscillation.

The frequency of the oscillating state is higher in LCDR than in LMA. A perturbative analysis of the LV equations supports this observation (appendix D). In fact, its prediction agrees well for a range of values of $\alpha$ (Fig.~\ref{fig:LV}D). This analysis also predicts that the oscillation in LV predator-prey model is extremely sensitive to concentration dependence of rates: even weak concentration dependence is sufficient to break the robustness of the oscillation. This result indicates that very special conditions are required for oscillatory reactions to remain stable, such as finite intrinsic reaction rate, $\kappa_I$, or presence of intrinsic or extrinsic noise.

\subsubsection{Stability of steady states in the presence of finite intrinsic reaction rates}

\begin{figure}
  \centering
  \includegraphics[width=0.9\columnwidth]{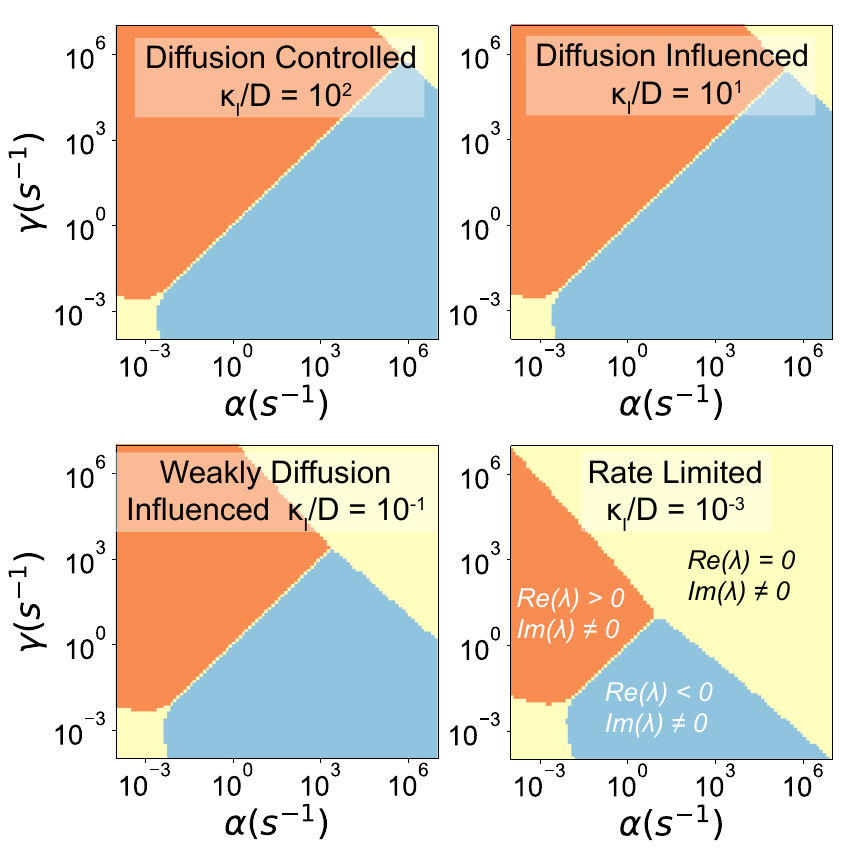}
  \caption{Phase diagram of LV oscillators when the intrinsic reaction rate is finite for different intrinsic rates $\kappa_I$ at $D = 1\mu m^2/s$. We have used the classification scheme provided in~\cite{yogurtcuTheoryBimolecularAssociation2015} to designate different regimes. We observe that fine-tuned oscillation persists even when the reactions are limited by the intrinsic reaction rates, although the corresponding parameter range narrows as $\kappa_I$ is decreased. These results suggest that effect of diffusion may be felt even when the reaction rate is controlled almost entirely by the intrinsic rate $\kappa_I$.}
  \label{fig:LV_pd_kI}
\end{figure}

So far, we have only considered the effect of the concentration dependent reaction rate in the diffusion limited regime. In this regime, the intrinsic reaction rate, $\kappa_I \rightarrow \infty$. However, in biological or physicochemical systems, we often encounter situations in which $\kappa_I$ is finite. Under such circumstances, the reaction kinetics may be entirely determined by the intrinsic reaction rates, which in turn, will restore the robust oscillatory behavior of the LV model. Therefore, we sought to understand how finite $\kappa_I$ modifies the phase diagram obtained in the diffusion limited regime (Fig.~\ref{fig:LV}C). To do so, we define the effective predation rate $\beta_{eff}$:
\begin{equation}
\beta_{eff} = \left( \frac{1}{\kappa_I} + \frac{1}{\beta(x,y)}\right)^{-1}, 
\end{equation}
where $\beta(x,y)$ is the concentration dependent encounter rate. We use this effective predation rate to calculate the steady state solutions of the LV equations and the stability of the solutions. 

Following the classification described in \cite{yogurtcuTheoryBimolecularAssociation2015}, we investigated four different $\kappa_I$ values for which the transient kinetics are diffusion controlled ($100 \mu m^2/s$), diffusion influenced ($10 \mu m^2/s$), weakly diffusion influenced ($0.1 \mu m^2/s$), and intrinsic reaction rate limited ($0.001 \mu m^2/s$). We find that fine-tuned oscillation persists in all four cases, even when the transient kinetics is rate limited and diffusion plays a negligible role (Fig.~\ref{fig:LV_pd_kI}). It happens because the stability of the steady state behavior is not only determined by the effective predation rate, but also its derivative at the steady state (see appendix D). The derivative has non-negligible, albeit diminishing, influence even when $\kappa_I << D$, which leads to fine-tuned oscillation even when $\beta_{eff}$ is essentially concentration independent. Because of the diminishing influence of the derivative, the parameter space for fine-tuned oscillation shrinks with decreasing $\kappa_I$. This result suggests that the influence of the diffusive transport process can impact steady state behavior even when it has negligible impact on the transient kinetics. In particular, in the context of our results, we find that fine-tuned oscillation is a robust behavior that persists even when the intrinsic reaction rates are several orders of magnitude smaller than the diffusion coefficient.

\subsubsection{Intrinsic noise leads to the emergence of concentration dependent rates}
\begin{figure}
  \centering
  \includegraphics[clip,width=0.7\columnwidth]{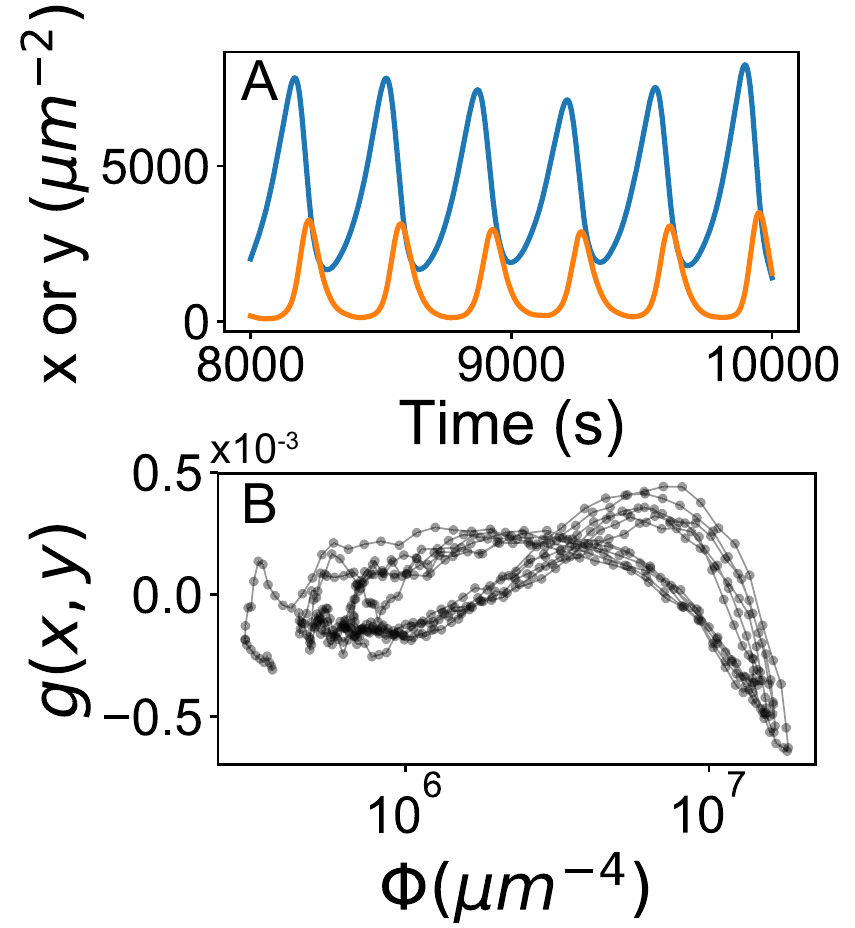}
  \caption{\textbf{Emergent concentration dependent rates in the stochastic Lotka-Volterra equation:} (A) A typical stochatic trajectory for $\alpha = 0.01 s^{-1}$, $\beta_0 = 0.00001 \mu m^{2}s^{-1}$,  and $\gamma = 0.04 s^{-1}$, $x(t=0) = y(t=0) = 1000 \mu m^{-2}$. The blue line represents prey concentration ($x$) and the orange line represents predator concentration ($y$). (B) For this trajectory we measured the covariance per unit $\Phi$, $g(x,y) = G(x,y)/\Phi$ vs $\Phi$. $g(x,y)$ is the equivalent of $\kappa(\Phi) - \kappa_0$ and it has a nonlinear dependence on the mass action $\Phi$, which suggests that intrinsic noise can lead to concentration dependent reaction rates.}
  \label{fig:LV-stoch}
\end{figure}

The presence of intrinsic or extrinsic noise is an important determinant of the steady state behavior, which we have not considered so far in this analysis. It is well-known that noise can significantly change the behavior of chemical oscillators~\cite{giverEffectsIntrinsicFluctuations2013,goldsteinSynchronizationPatternsGeometrically2015,mckaneStochasticPatternFormation2014,duncanNoiseinducedMultistabilityChemical2015,biancalaniGiantAmplificationNoise2017}. Its impact on the canonical LV systems, systems without concentration dependent rates, have been widely studied. In particular, applying a chemical master equation for the birth-death processes to the LV reactions (Eq.~\ref{Eq:LV-reactions}), it can be shown~\cite{curtisDynamicsPredatorPreyModel2017} that in the presence of intrinsic noise the mean concentrations of the predator and the prey follow modified deterministic equations:
\begin{eqnarray}
\frac{dx}{dt} &=& \alpha x - \beta_0xy -\beta_0G(x,y)\nonumber\\
\frac{dy}{dt} &=& \beta_0xy+\beta_0G(x,y) - \gamma y,
\label{Eq:LV-SDE}
\end{eqnarray}
where $G(x(t),y(t)) = cov(x(t),y(t))$ is the (time-dependent) covariance between the prey and the predator population. This equation has the same form as Eq.~\ref{Eq:LV-ODE}, with $\beta_0xy  + \beta_0 G(x,y)$ acting as the concentration dependent predation propensity, $\beta(x,y)xy$.  In particular, we find that for typical stochastic trajectories obtained from the canonical LV model (Fig.~\ref{fig:LV-stoch} A), $G(x,y)$ has nonlinear dependence on $\Phi$. To see this, we define $g(x,y) = G(x,y)/\Phi$, which plays the same role in this context as $\kappa(\Phi) - \kappa_0$ does in the context of diffusion-induced concentration dependence. As Fig.~\ref{fig:LV-stoch}B shows, $g(x,y)$ has a complex dependence on the mass action $\Phi$, which suggests that intrinsic noise may lead to the emergence of concentration dependent reaction rates. 

We must stress that the concentration dependence of $g(x,y)$ arose even when the predation rate $\beta_0$ was concentration independent. That is, the emergence of concentration dependent predation rate is not limited to dimensions less than three, and we can expect to observe its effect even in 3D.  Moreover, even though the functional form of $g(x,y)$ is very different than $\beta(x,y)$, we expect the same drastic shrinkage of the parameter space available for stable oscillation of the concentrations, since presence of any concentration dependence leads to fine-tuning of the oscillation (appendix D). This observation may explain why it is difficult to obtain stable oscillation in stochastic LV predator-prey model. However, more careful studies are needed to establish this claim.  

Finally, here we have shown the effect of intrinsic noise only for cases when $\beta$ is not concentration dependent. It will be interesting to study the combined impact of the concentration dependent predation rate $\beta(x,y)$ and the emergent rate $g(x,y)$ on the steady state stability of LV model and other dynamical systems. Moreover, it is likely that similar concentration dependent rates will arise in biologically relevant situations also, which are noisy chemical systems. Therefore, a careful classification of the biological systems in terms of concentration dependence of the reaction rates is necessary. A classification scheme, in terms of accuracy of well-mixed reactions, has been suggested in ref~\cite{yogurtcuTheoryBimolecularAssociation2015}. However, as we have shown, we also need to consider the impact of diffusion and noise on such a classification. 

\section{Discussion}

\paragraph*{Concentration dependent reaction rate} In this paper, we have used a hierarchical multiscale simulation framework to identify the origin of concentration dependence of the rate of association reactions and its consequences. We find that diffusion in 2D leads to the observed concentration dependence, and this dependence is universal across the different types of association reactions and the pair potentials considered here. In particular, the concentration dependent rate $\kappa(\Phi)$ can be written in a simple empirical form, which leads to drastic changes in the steady state properties of model chemical systems.

$\kappa(\Phi)$ is characterized by four parameters $\kappa_0$, $\kappa_{max}$, $\epsilon$, and $\Phi_0$. When $\Phi \rightarrow 0$, $\kappa(\Phi)\rightarrow \kappa_0$, which is a concentration independent constant. $\kappa_0$ depends on the interaction potential, the reaction radius, and the diffusion coefficient. However, the diffusion coefficient merely rescales the value of $\kappa_0$ (Fig.~\ref{fig:SI-D-dependence}), whereas the interaction potential and the reaction radius has more substantial impact. $\kappa_0$ depends on the interaction potential: $\kappa_0$ is higher for attractive LJ interaction than repulsive WCA interaction (Fig.~\ref{fig:kappa}B) for otherwise identical systems. Furthermore, we find that the lower the reaction radius the lower the value of $\kappa_0$, which is what we expect (Fig.~\ref{fig:kappa}B). In particular, when $r_a = 0.89r_{min} \approx \sigma$, the value of $\kappa_0$ is comparable to its value computed using Smoluchowski's theory of diffusion limited association reaction, which uses purely collision-based interaction between molecules~\cite{yogurtcuTheoryBimolecularAssociation2015, dibakDiffusioninfluencedReactionRates2019} (appendix E). For all the analysis involving $\kappa(\Phi)$, we have assumed that $D = 1\mu m^2/s$ and $r_a = 1.2 r_{min}$, such that all $\kappa_0$ values are around $4 \mu m^2/s$. While the empirical laws fitted in Fig.~\ref{fig:empirical}B do depend on the value of $\kappa_0$, it does not make it less general. In fact, computation expense permitted, we can repeat the same set of analysis using $r_a = 0.89 r_{min}$, which will have a different $\kappa_0$ and $\Phi_0$ values, but same values of $\kappa_{max}$ and $\epsilon$. 

$\kappa_{max}$ is the value of $\kappa(\Phi)$ in the other extreme, when $\Phi \rightarrow \infty$. As we have discussed in section II.C, the value of $\kappa_{max}$ can be measured by running simulations at higher packing fractions of the molecules, which we will do in a future paper. At higher packing fractions, the diffusive encounter rate is affected by the dynamic variation of the viscosity of the reactant solution. The variation of the viscosity depends on the interaction potential~\cite{pednekarSimulationShearThickening2017, bulowDynamicClusterFormation2019}. Therefore, we believe, the choice of attractive or repulsive interaction may lead to different $\kappa_{max}$ values. However, in this paper, we ignore this difference. Instead, irrespective of the interaction potential, we have used a meanfield formula derived using Smoluchowski's theory to estimate the value of $\kappa_{max}$, which is around $20 \mu m^2/s$ (appendix E). 

Between $\kappa_0$ and $\kappa_{max}$, $\kappa(\Phi)$ has a sigmoidal variation that is characterized by an exponent $\epsilon$ and a mass action scale $\Phi_0$. Through data analysis, we find that $\epsilon$ is equal to $3/8 = 0.375$. While we cannot provide a fundamental reason behind this exponent, the meanfield theory (appendix E) predicts that for low $\Phi$ values $\kappa(\Phi) - \kappa_0$ indeed scales as $\Phi^{0.4}$, which is remarkably close to what we have found through our analysis. Therefore, it is likely that the $\Phi^{\epsilon}$ scaling emerges due to the properties of diffusion in 2D. 

The scale $\Phi_0$, in effect, measures the contribution of the concentration dependent rates on the reaction kinetics. If $\Phi$ values are much lower than $\Phi_0$, the concentration dependent rate is subdominant and the behavior of the chemical system in this regime is well approximated by LMA. However, for $\Phi$ values near or higher than $\Phi_0$, the reaction rates are strongly concentration dependent until they reach $\kappa_{max}$, beyond which the rates become concentration independent again. Crucially, $\Phi_0$ is system dependent and higher values of $\Phi_0$ imply broader range of validity of LMA. For example, for the heterodimerization reaction, i.e., $A + B \rightarrow AB$, $\Phi_0$ value is higher than that for the homodimerization reaction. Therefore, we expect LMA to describe association reactions between unlike species much better than between like species.

We expect $\Phi_0$ to depend nontrivially on the form of the interaction potentials. In this paper, we have only considered isotropic interactions of identical strength and measured their $\Phi_0$ values by fitting the functional form of $\kappa(\Phi)$ to the observed data. However, real systems often interact with each other through anisotropic interactions of varying strength. Understanding how $\kappa(\Phi)$ and $\Phi_0$ behaves in these systems will help establish the range of validity of LMA and render chemical kinetic models more accurate. In particular, it may be possible to estimate the values of $\kappa_0$ and $\Phi_0$ using self-consistent theories of association reaction in 2D~\cite{szaboFirstPassageTime1980, szaboTheoryDiffusioninfluencedFluorescence1989, torneyDiffusionLimitedReactionRate1983, yogurtcuTheoryBimolecularAssociation2015, gopichMultisiteReversibleAssociation2020}. 

\paragraph*{Dimensionality of concentration dependence} In this paper, the concentration dependence of the reaction rates stems from the peculiarities of diffusion in 2D. The reentrant nature of diffusion in dimensions lower than three leads to the observed concentration dependence, which, as we have exposed, leads to drastic changes in the steady state properties of dynamical systems. However, concentration dependence can originate from other factors as well, which we have not considered here. For example, viscosity can change dynamically at molecular packing fractions that are quintessentially found inside a cell. Such dynamic changes in viscosity can dramatically reduce both translational and rotational diffusion coefficients~\cite{bulowDynamicClusterFormation2019}, which can give rise to another form of concentration dependence of the diffusion limited reaction rates. Such crowding induced concentration dependence is not limited by the dimension of the problem and can be observed even in 3D~\cite{grimaSystematicInvestigationRate2006}. Furthermore, as we have shown, albeit non-rigorously, intrinsic noise can also lead to emergent concentration dependence of the association reaction rates, even when the original reaction rates are concentration independent constants. It is well known that diffusion limited reaction rates are concentration independent constants only for dimensions greater or equal to three. Therefore, intrinsic noise offers another mechanism to obtain concentration dependent reaction rates in 3D. We expect that these three different flavors of concentration dependence will have different impact on the transient kinetics and may even have different steady state behaviors. However, all of these claims warrant rigorous examination, which we hope will inspire many future investigations of the origin and the consequences of the concentration dependent rates.  

\paragraph*{Impact on biomolecular systems} Concentration dependence of the reaction rates leads to drastic changes in the behavior of simple chemical systems, which may have serious repercussions on the behavior of more complex chemical systems, such as the cell-signaling or metabolic reactions. We find that concentration dependent reaction rates lead to reduction in intrinsic noise in dimerization reaction. Crucially, they destabilize robust oscillation in the LV predator-prey system and renders it fine-tuned to the parameter values. Although LV model is less relevant for biomolecular systems, the results derived from this model system can be directly applied to study feedback driven oscillatory systems on cell membranes or other 2D surfaces inside the cell. For example, signaling circuits in cell growth and development do show oscillatory behavior~\cite{brandmanFeedbackLoopsShape2008, yanCalciumParticipatesFeedback2009, alonIntroductionSystemsBiology2019, palssonSystemsBiologyProperties2006, karrWholeCellComputationalModel2012}. These circuits are usually membrane bound. Therefore, we expect our results to be directly applicable to such systems. 

The concentration dependent rates have another important implication in the context of modeling of biomolecular systems. The systems biology models used to study complex biological functions, such as cell signaling or metabolism, often use reaction rates measured from experiments. More often than not, the measured reaction rates vary wildly over several orders of magnitude~\cite{gutenkunstUniversallySloppyParameter2007}. Although such variations are often attributed to experimental imprecision, our results offer a plausible alternate hypothesis. As we have seen, the concentration dependent rates can vary across several orders of magnitude. Also, it is well-known that cells modulate the concentration of the biomolecules to achieve different tasks. It is possible that the rates measured by different experiments had different concentration of the reactant molecules, which led to the broad variation of the measured rates. This hypothesis has two implications. Firstly, it implies that the larger the range of reaction rates, the larger is the concentration fluctuation of the reacting molecules. Secondly, it will be possible to fit a sigmoidal $\kappa(\Phi)$ using the measured rates and use that concentration dependence to construct a class of systems biology models. Given the concentration dependent reaction rates have such drastic impact on the simple systems, it will lead to very interesting and potentially undiscovered phenomena in complex chemical systems.

\paragraph*{Acknowledgement} This work was funded by an LDRD grant (XX01) from LANL and has the following LA-UR number:LA-UR-20-24037. Computations used resources provided by the LANL Institutional Computing Program, which is supported by the U.S. DOE National Nuclear Security Administration under Contract No. DE-AC52-06NA25396. The author thanks Prof. Angel E. Garcia, Dr. Van A. Ngo, Prof. Yair Shokef, and Dr. Sumit Majumder for useful discussion and critical reading of the manuscript. The author also thanks two anonymous referees whose inputs have significantly improved the quality of this paper. 

\appendix

\section{Glossary}
\begin{tabular}{l l}
  BD: & Brownian Dynamics\\
  GFRD: & Green's function reaction dynamics\\
  LCDR: & Law of concentration dependent rates \\
  LMA: & Law of mass action \\
  LJ: & Lennard-Jones \\
  LV: & Lotka-Volterra \\
  WCA: & Weeks-Chandler-Andersen\\
  $[X]$: & Concentration of a molecule $X$\\
  $\Phi$: & Mass action. Product of the concentrations\\ 
  & of the reactants \\
  $\kappa(\Phi)$: & Concentration dependent diffusion-limited \\
  &rate of reaction\\
  $r$: & Propensity of a reaction. $r = \kappa(\Phi)\Phi$\\
  $\mbox{nD}$: & n-dimension \\ 
\end{tabular}

%%%%%%%%%%%%%%%%%%%%%%

\section{Methods}
\subsection{Molecular Simulation}

We use a recently developed molecular simulation technique, called Green's function reaction dynamics or GFRD~\cite{takahashiSpatiotemporalCorrelationsCan2010,vijaykumarCombiningMolecularDynamics2015,vijaykumarMultiscaleSimulationsAnisotropic2017,sbailoEfficientMultiscaleGreen2017,sokolowskiEGFRDAllDimensions2019} to perform molecular simulation.

The basic tenet of GFRD is that tagged-particles, such as proteins, in biomolecular systems can be partitioned into two groups: (a) isolated particles, which freely diffuse and (b) interacting particles, which interact with other particles. In a special form of GFRD, the isolated particles are propagated in an event-driven manner using the Green's function of diffusion until the particle encounters another particle and it can no longer be treated as an isolated particle, whereas the interacting particles are propagated using a molecular mechanics algorithm, such as molecular dynamics, dissipative particle dynamics or Markov state models~\cite{vijaykumarCombiningMolecularDynamics2015}. In the present work, we use overdamped Brownian dynamics (BD) as our molecular mechanics algorithm. This updated form is called BD-GFRD~\cite{vijaykumarCombiningMolecularDynamics2015,vijaykumarMultiscaleSimulationsAnisotropic2017}.

We assumed that the particles of type $A$ are spheres of radius 2 nm and particles of type $B$ are of radius 2.9 nm (Fig.~\ref{fig:setup}B) and they interact with each other through short-ranged isotropic interactions, such as the 6-12 Lennard-Jones or WCA interaction, (Fig.~\ref{fig:setup}C) of strength $6k_BT$, where $k_B$ is the Boltzmann's constant and $T = 310 K$ is the temperature of the heat bath. In addition, the diffusion costants of particles of type $A$ and type $B$ were $1 \mu m^2/s$ and $0.69 \mu m^2/s$, respectively. Because we assume overdamped dynamics, their mass is irrelevant for the simulation. These parameter values correspond to typical protein-protein interaction parameters, e.g. Ras-Raf interaction~\cite{sarkarPresenceAbsenceRas2020}. Furthermore, we assume that the reactions happen on a 2D surface, such as the plasma membrane of a cell. Hence, all of our BD-GFRD simulations were done on a $1\mu m \times 1 \mu m$ plane with periodic boundary conditions. Using this setup we ran the simulations for 96 cpu-hours or $10^4$ dimerization events, whichever occured first. We could reach a timescale of about 100 ms for $300 /\mu m^2$, the largest concentration considered here.

While using spherical particles is necessary to use current BD-GFRD framework, the interaction potential need not be isotropic. However, for the ease of exposition we have used isotropic potential in this paper. We have studied reaction kinetics under both attractive and repulsive potential. We have modeled the former using a LJ potential with cutoff at $2.5\sigma$ and the latter with a WCA interaction~\cite{weeksRoleRepulsiveForces1971} (Fig.~\ref{fig:setup}C). In this paper we report results for cases when the interaction strength for both potentials were $6k_BT$. However, we have checked our results for various interaction strengths up to $10 k_BT$ and have found that variation of interaction energies within this range does not affect our results. The lengthscale of the interaction potentials, $\sigma$, was chosen to be equal to the sum of the radius of the two reactants.

We consider two different types of association reactions: homodimerization and heterodimerization. For the former, we consider reactions of type $A+A \rightarrow A_2$, while for the latter, we consider reactions of the form $A+B \rightarrow AB$. To simulate association reaction we used a special case of Doi's volume reaction model~\cite{doiTheoryDiffusioncontrolledReaction1975,doiTheoryDiffusioncontrolledReaction1975a}. We assume that the particles react as soon as their distance becomes less than or equal to the reaction distance $r_a$ (Fig.~\ref{fig:setup}D). The reaction radius can be larger or smaller than $r_{min}$, the location of the minimum of the interaction potential. Based on our previous work, unless otherwise stated, we assume that $r_a = 1.2r_{min}$. We have found that the concentration dependence does not depend qualitatively on $r_a$, but it has quantitative impact, which we have discussed in this paper.

\subsection{Measurement of $\kappa(\Phi)$}

We measured the reaction rates in an ensemble, in which, the total number of particles of all species remained constant. We started with all monomers and no homo- or heterodimers. Hence, we removed a product molecule from the simulation as soon as it formed and replaced it with the reactant molecules that it was formed from. We placed the reactant molecules randomly on the simulation box to avoid introduction of unwanted correlation. Because we consider only dilute systems, such random replacements very rarely encounter another molecule after the replacement and, hence, does not break detailed balance. Furthermore, if there is such an encounter, we reject the random replacement and repeat the random placement until there is no encounter. In fact, this process is equivalent to starting with a new initial condition after each dimerization event. For denser systems, which we have not considered in this paper, we have to use sampling techniques, such as the metropolis algorithm, that preserve detailed balance.

To measure the reaction rates, we computed the probability distribution function (PDF) of the time interval between two product formation reactions. The reaction propensities, $r$, were then calculated by fitting the exponential tail of the PDFs. The exponential tail results from the reactions, whose rates are lower than the diffusive encounter rate. Therefore, we can use well-mixed approximation, i.e., Eq.~\ref{eqn:kappa_phi} to calculate $\kappa(\Phi)$.

\begin{equation}
  r = \kappa(\Phi)\Phi
  \label{eqn:kappa_phi_app}
\end{equation}

\subsection{Chemical kinetic model}

We use $\kappa(\Phi)$ to construct chemical kinetic models that were solved using Gillespie algorithm or analytical calculation. For the dimerization reactions, each stochastic simulation were run until the simulation time reached $t_{max} = 100 s$ or $100/k_{off}$, whichever was shorter. The steady state values were measured by averaging the observables between $t_{max}/2$ and $t_{max}$ over 100 different replicates. The system reached steady state within $t_{max}/10$ for all parameter values explored. The chemical kinetic equation for the product concentration, $P$, is given by:
\begin{eqnarray}
  \frac{dP}{dt} &=& k_{on} - k_{off}[P]\\
 % &=& \kappa(\Phi)\Phi - \kappa(\Phi)k_{D}[P]
\end{eqnarray}
In this equation, $k_{on} = r(\Phi) = \kappa(\Phi)\Phi$ is the propensity of the concentration dependent diffusion-influenced reaction and $\Phi$ is the mass action of the reaction. For LMA, $k_{on}$ is $\Phi$-independent constant, whereas for LCDR, $k_{on}$ depends on $\Phi$. $k_{off}$ is the dissociation rate. For reversible reactions approaching chemical equilibrium, $k_{off}$ is given by the product of the association rate $k_{on}$ and the equilibrium constant $k_{D}$, which ensures detailed balence. For nonequilibrium steady states, $k_{off}$ were varied independently from $k_{on}$. 

%%%%%%%%%%%%%%%%%%%%%%%%%%%%%%%%%%%%%%%%%%%%%

\section{Steady state concentration of molecules in dimerization reaction}
\subsection{Equation for steady state concentration}

Let's consider the following dimerization reaction.
\begin{equation}
A + A \underset{k_{off}}{\stackrel{k_{on}}{\rightleftharpoons}} A_2
\end{equation}
The concentrations of the molecules obey the following set of kinetic equations.
\begin{eqnarray}
  \frac{d[A]}{dt} &=& -2k_{on}\Phi + 2k_{off} [A_2]\\
  \frac{d[A_2]}{dt} &=& k_{on}\Phi - k_{off}[A_2]\\
  \Phi &=& \frac{[A]([A]-1)}{2} \\
  k_{on} &=& \left\{\begin{array}{lr}
  k_0 & \text{for LMA} \\
  \kappa(\Phi)
   & \text{for LCDR}\end{array}\right. \\
    \kappa(\Phi) &=& 
  \frac{\kappa_{max}\kappa_0}{\kappa_0 + (\kappa_{max} - \kappa_0)\exp\left[-\left(\frac{\Phi}{\Phi_0}\right)^{\epsilon}\right]} 
  \label{eq:kappa_defn}
\end{eqnarray}
  
In steady state the following relation holds:
\begin{equation}
  k_{on}\Phi = k_{off}[A_2] \label{eqn:ss}.
\end{equation}
In addition, we assume that the total number of monomers, $N$, is fixed, such that:
\begin{equation}
  [A] + 2[A_2] = N \label{eqn:A_consv}
\end{equation}

Combining Eqns. \ref{eqn:ss} and \ref{eqn:A_consv}, we get the following equation, whose roots are the steady state concentrations of the monomers.
\begin{equation}
  k_{on}\times x(x-1) - k_{off}(N-x) = 0 \label{eqn:ssm},
\end{equation}
where $x = [A]$ is the steady state concentration of the monomer and $\Phi = x(x-1)/2$ is the mass action.

\subsection{Solution}

The solution of Eq.~\ref{eqn:ssm} depends on the nature of the steady state. For example, if the steady state is at chemical equilibrium, then $k_{off}$ has to be equal to $k_{on}k_D$ at all times, where $k_D$ is the equilibrium constant. As a result, for both LMA and LCDR, Eq.~\ref{eqn:ssm} simplifies to 
\begin{equation}
  x(x-1) - k_D(N-x) = 0 \label{eqn:ssm_eq},
\end{equation}
which has a simple solution: 
\begin{equation}
x_{Eq} = \frac{\sqrt{(k_D-1)^2 + 4k_DN}  - (k_D - 1)}{2}. \label{eqn:dim_MA_soln}
\end{equation}

However, when the steady state is a nonequilibrium steady state, $k_{off}$ does not have to be exactly equal to $k_{on}k_D$; we can control it independently from $k_{on}$. For the sake of simplicity let's assume that $k_{off} = k_D$. Furthermore, we assume that under LMA $k_{on} = k_0$, where $k_0$ is a constant, from which we find that the solution at the nonequilibrium steady state is:
\begin{eqnarray}
x_{MA} &=& \frac{\sqrt{(k_D-k_0)^2 + 4k_Dk_0 N}  - (k_D - k_0)}{2k_0}\\
&=& \frac{\sqrt{(k_D^{sc}- 1)^2 + 4k_D^{sc} N}  - (k_D^{sc} - 1)}{2},\label{eqn:dim_MA_soln}
\end{eqnarray}
where $k_D^{sc}$ is equal to $k_D/k_0$. It is evident from this solution that, under LMA, it is not possible to attain a nonequilibrium steady state by manipulating $k_{off}$. Doing so, merely scales $k_D$ to a different value and the steady state fluctuations are equilibrium fluctuations that obey detailed balance. 

\begin{figure}
  \centering
  \includegraphics[width=0.5\textwidth]{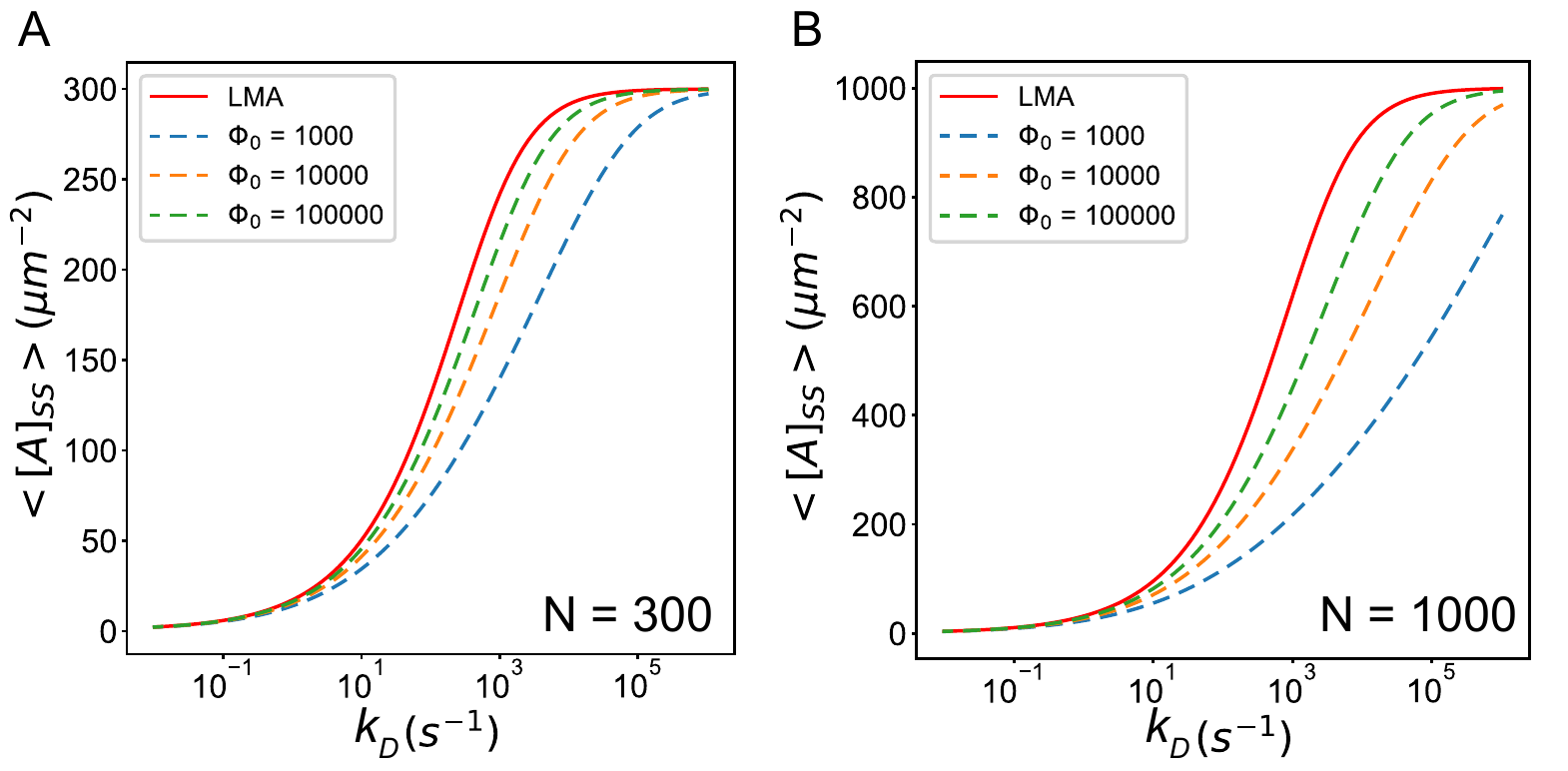}
  \caption{Variation of steady state monomer concentration with $k_D$. $N = 1000$, $\kappa_0 = 1$ and $\kappa_{max} = 1000$ were used for all plots. $\Phi_0$ values for LCDR cases are shown in legend. $k_0 = 1$ was used for the LMA solution.}
  \label{fig:SI_ssn}
\end{figure}

In contrast, when LCDR is used it is possible to break detailed balance by manipulating $k_{off}$. To show this we compute the steady state solution using $k_{on} = \kappa(\Phi)$ and $k_{off} = k_D$. Unfortunately, closed form solution is not easy to obtain. So, we use Newton-Raphson root finding algorithm to find the solutions of Eq.~\ref{eqn:ssm}. Steady state solutions for a few $\Phi_0$ and $N$ are shown in Fig.~\ref{fig:SI_ssn}. Clearly, the nonequilibrium steady state solutions are different for LCDR than the equilibrium solution that we obtain for both LMA and LCDR. In general, the higher the $\Phi_0$, the more similar are the LCDR and LMA solutions. Physically, it implies that the validity of LMA increases as the difficulty for an association reaction increases. If the association reaction is a rare event, then the reaction events will follow Poisson distribution. Hence, the reaction rate will be a concentration independent constant.

%%%%%%%%%%%%%%%%%%%%%%%%%%%%%%%%

\section{Steady state behavior of Lotka-Volterra model}
We study the Lotka-Volterra (LV) predator-prey equations~\cite{murray2007mathematical}, which describes the dynamics of the following set of reactions:
\begin{eqnarray}
X &\rightarrow& 2X\\
X + Y &\rightarrow& 2Y \\
Y &\rightarrow& \phi,
\end{eqnarray}
where $X$ is the prey and $Y$ is the predator, and $\phi$ represents death. The most general LV equation can be written in the following way:
\begin{eqnarray}\label{eqn:SI_LV}
\frac{dx}{dt} &=& \alpha x - \beta(x,y)xy\\
\frac{dy}{dt} &=& \beta(x,y)xy- \gamma y,
\end{eqnarray}
where $\alpha$ is the birth rate of the prey, $\beta(x,y)$ is a concentration dependent predation rate, and $\gamma$ is the death rate of the predator. The equation for the steady state concentrations are defined by the following two equations~\cite{strogatz2018nonlinear}:
\begin{eqnarray}
  y &=& \alpha/\beta(x,y)\label{eq:SI-LV-soln-x}\\
  x &=& \gamma/\beta(x,y)\label{eq:SI-LV-soln-y}.
\end{eqnarray}
The Jacobian of Eq.~\ref{eqn:SI_LV} is:
\begin{equation}
  J(x,y) = \begin{pmatrix}
    \alpha - \frac{\partial\beta}{\partial x}xy - \beta y & - \frac{\partial\beta}{\partial y}xy - \beta x \\
    \frac{\partial\beta}{\partial x}xy + \beta y & \frac{\partial\beta}{\partial y}xy + \beta x - \gamma
\end{pmatrix}
\end{equation}

\subsection{Steady state behavior under LMA}
When LMA is used $\beta(x,y) = \beta_0$. Therefore, the steady state solutions are:
\begin{eqnarray}
  y &=& \alpha/\beta_0\notag\\
  x &=& \gamma/\beta_0,
\end{eqnarray}
and the Jacobian at this steady state value is:
\begin{equation}
  J(x,y) = \begin{pmatrix}
    0 & - \gamma \\
    \alpha & 0.
\end{pmatrix}
\end{equation}
Therefore, for all $\alpha > 0$ and $\gamma > 0$, the eigenvalues are $\pm\iota\sqrt{\alpha\gamma}$, i.e., the steady state solution is oscillating and it is robust to perturbation in $\alpha$ and $\gamma$

\subsection{Steady state behavior under LCDR}

\subsubsection{General steady state solution}

When LCDR is used $\beta(x,y) = \kappa(\Phi)$, for $\Phi = xy$. $\kappa(\Phi)$ is defined in Eq.~\ref{eq:kappa_defn}. Eqns.~\ref{eq:SI-LV-soln-x} and \ref{eq:SI-LV-soln-y} can be used to write the following equation for $\Phi$.
\begin{equation}
  \Phi\beta^2(\Phi) = \alpha\gamma
\end{equation}
By dividing both sides with $\Phi_0$ and defining $z = \Phi/\Phi_0$, we write the reduced equation for the steady state solution.
\begin{equation}
  z\beta^2(z) - \frac{\alpha\gamma}{\Phi_0} = 0
\end{equation}
If $z_{ss}$ is the solution of this equation, then the steady state concentrations of prey and predator are:

\begin{eqnarray}\label{eq:LV-ss-soln}
  x_{ss} &=& \gamma/\beta(z_{ss})\notag\\
  y_{ss} &=& \alpha/\beta(z_{ss})
\end{eqnarray}

The Jacobian also can be easily constructed by noting that $\frac{\partial\beta}{\partial x \text{ or } y} = \frac{x\text{ or }y}{\Phi_0}\frac{\partial\beta}{\partial z}$.

\subsubsection{Perturbation theory}
We can gain some insights about the stability of the steady state solution by solving the equations~\ref{eq:SI-LV-soln-x} and \ref{eq:SI-LV-soln-y} using perturbation theory. To do so, we note that near $z=0$, $\beta(z)\approx \kappa_0 + \kappa_1 z^\epsilon \approx \beta_1\Phi^{\delta} = \beta_1(xy)^\delta$. In general, $\delta$ is a function of $\Phi$ and $\beta_1$ is a constant. In particular, $\delta \rightarrow 0$ as $\Phi \rightarrow 0$, and $\delta \rightarrow \epsilon$ for $\Phi >> \Phi_0$. Under this approximation the steady state solutions are:
\begin{eqnarray}
  \Phi_{ss} &=& \left(\frac{\alpha\gamma}{\beta_1^2}\right)^{\frac{1}{2\delta+1}}\\
  \beta(x_{ss},y_{ss}) &=& \beta_1^{1/1+2\delta}\times (\alpha\gamma)^{\delta/1+2\delta}\\
  x_{ss} &=& \frac{\gamma}{\beta(x_{ss},y_{ss})}\label{eqn:x_ss}\\
  y_{ss} &=& \frac{\alpha}{\beta(x_{ss},y_{ss})}.
\end{eqnarray}

In particular, when $\alpha = \gamma$, their values are: 

\begin{eqnarray}\label{eq:LV-ss-soln}
  x_{ss} = y_{ss} &=& \left(\frac{\gamma}{\beta_1}\right)^{\frac{1}{1+2\delta}},\label{eqn:xy_ss}
\end{eqnarray}
which will scale as $\gamma$, when $\Phi\rightarrow 0$, such that $\delta \rightarrow 0$, and will scales as $\gamma^{\frac{1}{1+2\epsilon}}$, when $\Phi \sim \Phi_0$ or higher. 

\begin{figure}
  \centering
  \includegraphics[width=0.8\columnwidth]{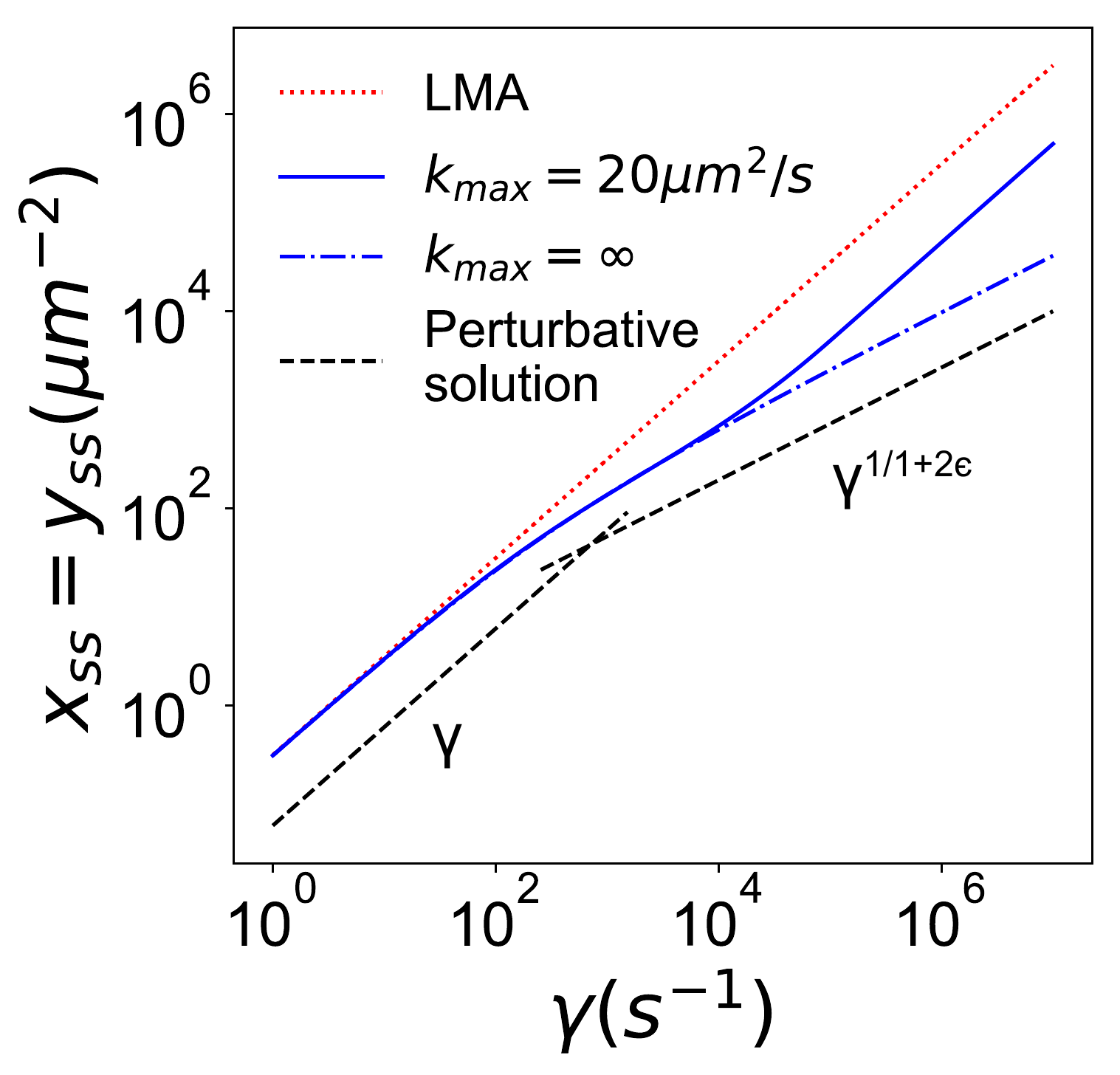}
  \caption{The steady state concentration of the oscillatory state is given by Eq.~\ref{eqn:xy_ss}, when $\gamma = \alpha$. The LMA steady state increases linearly with $\gamma$ (red dotted line), but the LCDR steady state increases sublinearly except for $\gamma \approx 0$ (blue solid line) for both $k_{max} = 20 \mu m^2/s$ and $k_{max} = \infty$. For $k_{max} = \infty$, such that $\kappa(\Phi)$ is always $\kappa_0 + \kappa_1 \Phi^{\epsilon}$, the LCDR steady state solution is correctly given by the perturbation theory (black dashed line), which predicts that $x_{ss}$ and $y_{ss}$ scale as $\gamma$, when $\Phi$, i.e. $x_{ss}$ is small and they scale as $\gamma^{1/1+2\epsilon}$, when $\Phi > \Phi_0 = 1\times10^5$.}
  \label{fig:PT_osc_ss}
\end{figure}

The Jacobian is:
\begin{equation}
  J(x,y) = \begin{pmatrix}
    \alpha - y\beta(x,y)(1+\delta) & - x\beta(x,y)(1+\delta) \\
    y\beta(x,y)(1+\delta) & x\beta(x,y)(1+\delta) - \gamma
\end{pmatrix},
\end{equation}
which in steady state has the simple form
\begin{equation}
  J(x_{ss},y_{ss}) = \begin{pmatrix}
    -\alpha\delta & -\gamma(1+\delta) \\
    \alpha(1+\delta) & \gamma\delta
\end{pmatrix}.
\end{equation}
The eigenvalues of this Jacobian satisfy the following characteristic equation.
\begin{eqnarray}
  (\lambda -\delta\gamma)(\lambda +\delta\alpha) + (1+\delta)^2\alpha\gamma &=& 0\\
  \Rightarrow \lambda^2 + \delta(\alpha -\gamma)\lambda + (1+2\delta)\alpha\lambda &=& 0
\end{eqnarray}
Therefore, the eigenvalues are:
\begin{equation}
  \lambda_{\pm} = -\frac{(\alpha -\gamma)\delta}{2} \pm \frac{\sqrt{\delta^2(\alpha-\gamma)^2 - 4(1+2\delta)\alpha\gamma}}{2}.
\end{equation}
 When $\delta = 0$, we recover the robust oscillatory behavior, as expected from LMA. However, as soon as $\delta > 0$, oscillation happens only when $\alpha = \gamma$! This is a drastic shrinkage of the parameter space compared to LMA. In fact, in LCDR, to get sustained oscillation, the parameters have to be fine-tuned such that $\alpha = \gamma$. Moreover, when oscillation happens, it happens with a frequency of $\sqrt{(1+2\delta)\alpha\gamma}$, which is $\sqrt{1+2\delta}$ times faster than the frequency of oscillation in LMA.

 When $\alpha > \gamma$, the real part of the eigenvalues are negative and the steady state is a stable fixed point. On the other hand, when $\gamma > \alpha$, the real part of the eigenvalues is greater than zero and the steady state is unstable to perturbations and the predator population grows exponentially and the prey populations shrinks exponentially, eventually resulting in extinction of the both prey and predator population. However, one should note that true collapse of the population is not possible in ODE based model, because $x = 0, y = 0$ is a saddle point, which means that the population can recover as long as it is greater than 0.

\begin{figure}
  \centering
  \includegraphics[width=\columnwidth]{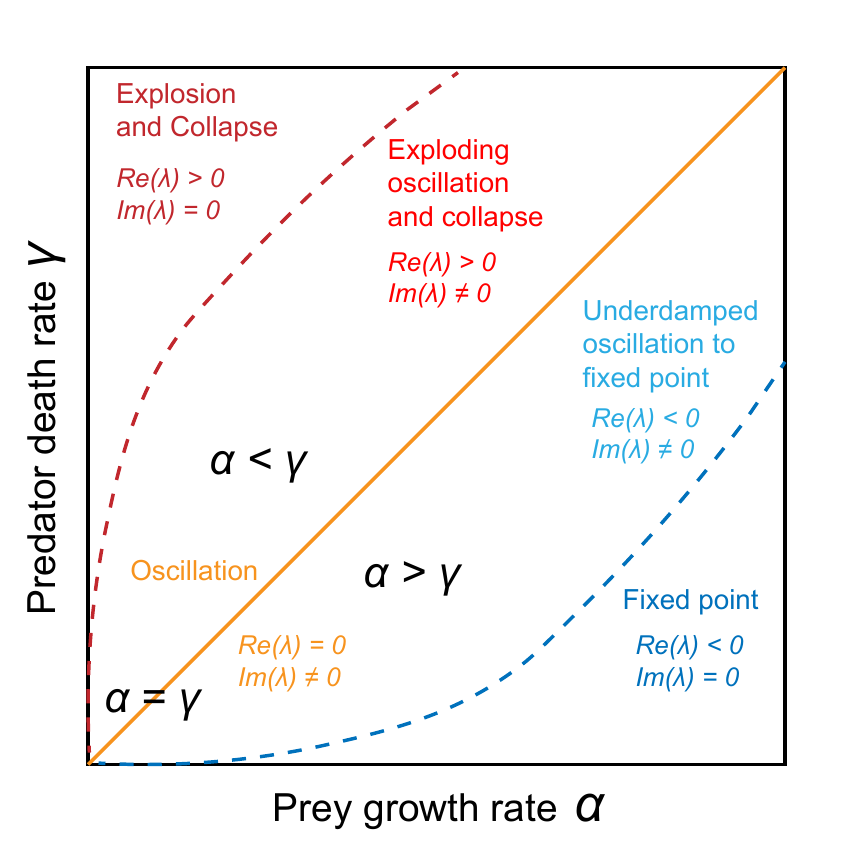}
  \caption{Phase diagram obtained using the perturbation theory. Perturbation theory correctly predicts the phase boundary between stable fixed points, unstable fixed points, and sustained oscillation ($\alpha = \gamma$, yellow line). However, it incorrectly predicts the phase boundaries that distinguish oscillating from non-oscillating approach to the fixed points (broken blue and crimson lines) at small $\alpha$ and $\gamma$.  A key drawback of the perturbation theory is that it does not account for the sigmoidal nature of $\kappa(\Phi)$ and its saturation at $\kappa_{max}$, which removes the non-oscillating regions (Explosion and Collapse \& Fixed point) from the phase diagram.}
  \label{fig:PT_phase_diagram}
\end{figure}

 When $\alpha \neq \gamma$, the population may reach its steady state through oscillation or without oscillation. The crossover between these two regions happen when the imaginary part of the eigenvalues become 0. That is when,
 \begin{eqnarray}
   \delta^2(\alpha-\gamma)^2 &=& 4(1+2\delta)\alpha\gamma \\
   \Rightarrow \gamma &=& \frac{\eta + 1}{\eta - 1}\alpha, \text{ when } \alpha < \gamma\\
   &=& \frac{\eta - 1}{\eta + 1}\alpha \text{ when } \alpha > \gamma\\
   \eta &=& \sqrt{\frac{(1+\delta)^2}{1+2\delta}}.
 \end{eqnarray}
The crossover region is defined by a line with slope $m = \frac{\eta \pm 1}{\eta \mp 1}$. However, we note that $\delta$ increases with $\Phi$. Hence, $m$ changes with $\Phi$. At large enough $\Phi$, $m \approx 1$, whereas when $\Phi$, hence $\delta$, is small, the slope  $m \rightarrow \infty$ or $\rightarrow 0$, depending on whether $\alpha < \gamma$ or $\alpha > \gamma$, respectively. We can combine the preceding analysis to find the steady state solution of the oscillatory state (Fig.~\ref{fig:PT_osc_ss}) and construct a phase diagram, which is shown in Fig.~\ref{fig:PT_phase_diagram}.

%%%%%%%%%%%%%%%%%%%%%%%%%%%%%%%%

\section{Computation of $\kappa_{max}$ using Smoluchowski theory}

\begin{figure}
  \centering
  \includegraphics[width=0.7\columnwidth]{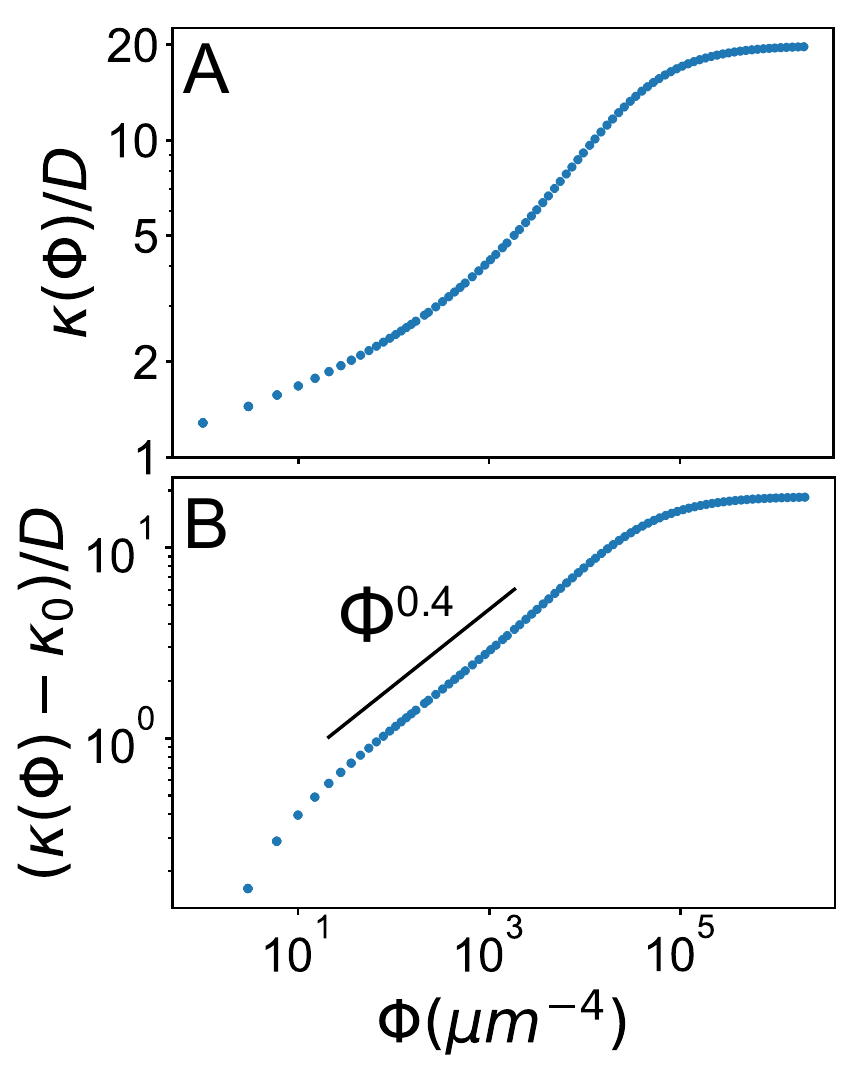}
  \caption{$\kappa(\Phi)$ obtained using Smoluchowski theory with meanfield approximation. (A) $\kappa(\Phi)$ vs $\Phi$ shows that $\kappa(\Phi)/D$ saturates at around $20$, which we take to be the value of $\kappa_{max}$. The value of $\kappa_0/D$ is between 1 and 2. (B) $\kappa(\Phi) - \kappa_0$ scales as $\Phi^{0.4}$ for small $\Phi$ values.}
  \label{fig:yogurt}
\end{figure}

The concentration dependent reaction rate in the presence of collisional-interaction is given by: 

\begin{align}
\kappa &= 8\pi D \left( \frac{4\log\frac{b}{\sigma}}{\left(1 - \sigma^2/b^2\right)^2} - \frac{2}{\left(1 - \sigma^2/b^2\right)} - 1\right) ^{-1}\label{eqn:kappa_rho}, 
\end{align}
where $D$ is the diffusion coefficient, $\sigma$ is the sum of the radius of two reacting molecules, and $b$ is the average radius of a circular region where only one reaction is possible~\cite{yogurtcuTheoryBimolecularAssociation2015}. Clearly $b$ is a concentration dependent quantity and the higher the concentration of the reactants the lower the value of $b$. In fact, we can estimate $b$ as a function of concentration using a meanfield approximation. If the mass action is $\Phi$ and the area of the simulation box is $A$, then on an average, the area per reaction is $A/\Phi$ and $b$ is given by~\cite{yogurtcuTheoryBimolecularAssociation2015}: 
\begin{align}
b(\Phi) = 2\sqrt{\frac{A}{\pi \Phi} - r_1^2 - r_2^2}, 
\end{align}
where $r_1$ and $r_2$ are the radius of the two interacting molecules. Plugging this expression in Eq.~\ref{eqn:kappa_rho} for $A_2$, we get figure~\ref{fig:yogurt}, which shows that $\kappa_{max}\approx 20$ and $\epsilon$ is approximately $0.4$ for small $\Phi$ values. Furthermore, $\kappa_0$ can be estimated by extrapolating Eq.~\ref{eqn:kappa_rho} to the $\Phi \to 0$ limit. Practically, it is estimated by finding the value of $\kappa/D$ at $\Phi = 1$, which is approximately $1.2$ (Fig.~\ref{fig:yogurt}A).

The value of the exponent derived from the meanfield theory is remarkably close to the values found through the analysis of the simulation data, but it is not exactly equal to what we found. This difference stems from the fact that  $A/\Phi$ underestimates the area available for each reaction. In reality, area available varies as $A/\Phi^{\alpha}$, where $\alpha < 1$. We found that as $\alpha$ is lowered from 1, $\epsilon$ computed using  Eq.~\ref{eqn:kappa_rho} also decreases. Therefore, for a suitable $\alpha$,  the value of $\epsilon$  will be $0.375$. However, we need careful investigation of the diffusion limited reactions to identify such an $\alpha$.

\clearpage

\bibliography{p12}

\end{document}